 [2003/12/12 5.2/AAS markup document class]%

\documentclass[iop]{emulateapj}

\usepackage{hyperref}
\usepackage{epsfig}
\usepackage{natbib}
\usepackage{graphicx}    
\usepackage{makeidx}
\makeindex                               
\usepackage{verbatim}                                   
\usepackage{amsmath,amsthm,amsfonts,amsopn,amssymb}     
\usepackage{longtable}
\usepackage{afterpage}
\usepackage{ulem}
\usepackage{epstopdf}  
\usepackage{multirow}
\usepackage{color}
\usepackage[left]{lineno}
\usepackage{amsmath}
\usepackage{bm}

\usepackage{chngpage}

\accepted{in ApJ}

\slugcomment{}
\shorttitle{Planet Hunters 7}
\shortauthors{Schmitt et al. (2014)}

\begin{document}
\title{Planet Hunters VII.  Discovery of a New Low-Mass, Low-Density Planet (PH3~c) Orbiting Kepler-289 with Mass Measurements of Two Additional Planets (PH3~b and d)\footnote{This publication has been made possible by the participation of more than 290,000 volunteers in the Planet Hunters project. Their contributions are individually acknowledged at http://www.planethunters.org/authors}}
\author{
Joseph R. Schmitt\altaffilmark{1},
Eric Agol\altaffilmark{2},
Katherine M. Deck\altaffilmark{3},
Leslie A. Rogers\altaffilmark{4,5},
J. Zachary Gazak\altaffilmark{6},
Debra A. Fischer\altaffilmark{1},
Ji Wang\altaffilmark{1},
Matthew J. Holman\altaffilmark{7},
Kian J. Jek\altaffilmark{8},
Charles Margossian\altaffilmark{1},
Mark R. Omohundro\altaffilmark{8},
Troy Winarski\altaffilmark{8},
John M. Brewer\altaffilmark{1},
Matthew J. Giguere\altaffilmark{1},
Chris Lintott\altaffilmark{9,10},
Stuart Lynn\altaffilmark{10},
Michael Parrish\altaffilmark{10},
Kevin Schawinski\altaffilmark{11},
Megan E. Schwamb\altaffilmark{12},
Robert Simpson\altaffilmark{9},
Arfon M. Smith\altaffilmark{10}
} 
\email{joseph.schmitt@yale.edu}

\altaffiltext{1}{Department of Astronomy, Yale University, New Haven, CT 06511 USA}
\altaffiltext{2}{Department of Astronomy, Box 351580, University of Washington, Seattle, WA 98195, USA}
\altaffiltext{3}{Department of Physics and Kavli Institute for Astrophysics and Space Research, Massachusetts Institute of Technology, 77 Massachusetts Ave., Cambridge, MA 02139}
\altaffiltext{4}{Department of Astronomy and Division of Geological and Planetary Sciences, California Institute of Technology, MC249-17, 1200 East California Boulevard, Pasadena, CA 91125, USA}
\altaffiltext{5}{Hubble Fellow}
\altaffiltext{6}{Institute for Astronomy, University of Hawai’i, 2680 Woodlawn Dr, Honolulu, HI 96822, USA}
\altaffiltext{7}{Harvard-Smithsonian Center for Astrophysics, 60 Garden St., Cambridge, MA 02138}
\altaffiltext{8}{Planet Hunter}
\altaffiltext{9}{Oxford Astrophysics, Denys Wilkinson Building, Keble Road, Oxford OX1 3RH}
\altaffiltext{10}{Adler Planetarium, 1300 S. Lake Shore Drive, Chicago, IL 60605, USA}
\altaffiltext{11}{Institute for Astronomy, Department of Physics, ETH Zurich, Wolfgang-Pauli-Strasse 16, CH-8093 Zurich, Switzerland}
\altaffiltext{12}{Institute of Astronomy and Astrophysics, Academia Sinica: 11F of Astronomy-Mathematics Building, National Taiwan University. No.1, Sec. 4, Roosevelt Rd, Taipei 10617, Taiwan}


\begin{abstract}

We report the discovery of one newly confirmed planet ($P=66.06$ days, $R_{\rm{P}}=2.68\pm0.17R_\oplus$) and mass determinations of two previously validated \textit{Kepler} planets, Kepler-289~b ($P=34.55$ days, $R_{\rm{P}}=2.15\pm0.10R_\oplus$) and Kepler-289-c ($P=125.85$ days, $R_{\rm{P}}=11.59\pm0.10R_\oplus$), through their transit timing variations (TTVs).  We also exclude the possibility that these three planets reside in a $1:2:4$ Laplace resonance.  The outer planet has very deep ($\sim1.3\%$), high signal-to-noise transits, which puts extremely tight constraints on its host star's stellar properties via Kepler's Third Law.  The star PH3 is a young ($\sim1$ Gyr as determined by isochrones and gyrochronology), Sun-like star with $M_*=1.08\pm0.02M_\odot$, $R_*=1.00\pm0.02R_\odot$, and $T_{\rm{eff}}=5990\pm38$ K.  The middle planet's large TTV amplitude ($\sim5$ hours) resulted either in non-detections or inaccurate detections in previous searches.  A strong chopping signal, a shorter period sinusoid in the TTVs, allows us to break the mass-eccentricity degeneracy and uniquely determine the masses of the inner, middle, and outer planets to be $M=7.3\pm6.8M_\oplus$, $4.0\pm0.9M_\oplus$, and $M=132\pm17M_\oplus$, which we designate PH3~b, c, and d, respectively.  Furthermore, the middle planet, PH3~c, has a relatively low density, $\rho=1.2\pm0.3$ g/cm$^3$ for a planet of its mass, requiring a substantial H/He atmosphere of $2.1^{+0.8}_{-0.3}\%$ by mass, and joins a growing population of low-mass, low-density planets.

\end{abstract}


\section{Introduction}
\label{intro}

Since the first planet discoveries in the 1990s \citep{Wolszczan1992,Mayor1995}, more than 1400 planets have been discovered, according to the Exoplanet Archive \citep{Wright2011}.  \textit{Kepler} was launched in 2009 and obtained precise photometric measurements for $\sim160,000$ stars with nearly continuous coverage for four years.  More than 3500 candidates have been discovered via the planet transit method \citep{Borucki2011,Batalha2013,Burke2014}, with nearly 1,000 of them now confirmed \citep[e.g.,][]{Lissauer2014,Rowe2014}. The transit planet search (TPS) algorithm \citep{Jenkins2002,Jenkins2010} has been used to search \textit{Kepler} stellar light curves for the characteristic dips in flux indicative of a planet.  Those which meet a certain significance threshold are placed on the Threshold Crossing Event (TCE) list \citep{Tenenbaum2013,Tenenbaum2014}. The TCEs are further examined and can be upgraded to \textit{Kepler Object of Interest} (KOI) candidate status or downgraded as false positives (FPs), which may be erroneous signals or real, astrophysical phenomena like eclipsing binaries \citep{Prsa2011,Slawson2011}, either on the target star or in the background.

The photometric transit technique can determine the radius of a planet, but generally not the mass and hence does not immediately indicate if a transit signal is due to a planet or a binary star system.  In certain cases, these transiting planet candidates can be confirmed as planets through a statistical elimination of astrophysical false positives, such as eclipsing binaries \citep{Fressin2012}.  Validation by multiplicity has been used on hundreds of candidates in multiple planet candidate systems \citep{Lissauer2014,Rowe2014}.  This method uses the fact that there are very few false positives in multiple transiting planet candidate systems due to the rare occurrence of two independent, unlikely events (either multiple false positives or a false positive and a true planet) both occurring for the same star.  However, neither of these methods can determine the planetary masses or orbital parameters, which is necessary for further characterization of the planets.  To date, the only way to determine masses purely from a photometric light curve is through the detection and inversion of TTVs, deviations from a perfectly linear ephemeris due to gravitational interactions of neighboring planets. (see Section~\ref{sec:ttv}).  In this paper, strong TTVs of the planets orbiting PH3 allow for the determination of planetary masses and thus their confirmation as planets.

\subsection{Planet Hunters}

The Planet Hunters (PH) project\footnote{http://www.planethunters.org/} is one of the many projects within the Zooniverse\footnote{https://www.zooniverse.org/} citizen science program \citep{Lintott2008,Lintott2011,Fortson2012}.  The goal of this project is to assist in the analysis of \textit{Kepler} lightcurves by complementing the TPS algorithmic search with human eyes.  PH breaks \textit{Kepler} lightcurves into 30 day increments and asks users to mark transit-like signals.  Additionally, the Talk\footnote{http://talk.planethunters.org} discussion page allows users to interact with the science team and each other, discuss lightcurves, and even collect certain categories of potential phenomena, such as eclipsing binaries,  heartbeat stars, variable stars, microlensing, and circumbinary planets.  Since December, 2010, more than 22 million \textit{Kepler} 30-day increment lightcurves have been examined by more than 290,000 public volunteers, amounting to 180 years of 40-hour work weeks.

PH has detected more than 60 new planet candidates, including a large number in their host star's habitable zone (HZ)	 \citep{Fischer2012,Lintott2013,Wang2013a,Schmitt2014}, and two newly confirmed planets \citep{Schwamb2013,Wang2013a}.  The first known \textit{Kepler} seven planet candidate \citep{Cabrera2014} was also independently discovered by PH and orbits in a compact, solar system-like arrangement \citep{Schmitt2014}.  PH's main contribution to the population of \textit{Kepler} candidates has been in long period candidates.  This is due to the fact that PH users detect one transit at a time, making PH sensitive to even one and two transit systems, whereas computer algorithms typically require three or more transits and build up signal with an increasing number of transits, thus decreasing sensitivity to longer periods.  However, computer algorithms are more robust at identifying smaller planets with transits that are hard to detect by eye.  As such, most PH candidates are approximately Neptune-sized or larger, which is where PH approaches completeness for short periods \citep{Schwamb2012}.  

\subsection{KOI-1353}

KIC 7303287 was first found to have a planet candidate (KOI-1353.01) in \citet{Borucki2011}, an $18.6\pm5.3R_{\oplus}$ planet in a 125.87 day orbit.  Later, a 34.54 day planet candidate (KOI-1353.02) with a radius of $3.8\pm1.1R_{\oplus}$ was also discovered \citep{Batalha2013}.  PH users then noticed an additional 66 day planet candidate, which was later detected as a 330 day TCE with a loosely constrained radius of $4.23\pm3.98$ \citep{Tenenbaum2014}, an alias of the true period.  PH volunteers also discovered its five hour amplitude TTVs in the transit data and brought it to the attention of the science team via the Talk page (see Section~\ref{sec:agol}).  Closer examination of KOI-1353.01 showed that outer planet also exhibits TTVs, but with an amplitude of order minutes.  In addition to a sinusoidal component with period equal to the super period, we also can identify a chopping signal in the residuals of the middle planet, allowing us to break the mass-eccentricity degeneracy.  As described below in Section~\ref{sec:agol}, the inner planet is also confirmed by having a mass upper limit well within the planetary regime.  Therefore, KOI-1353 is confirmed as PH3~b, c, and d.  While in preparation of this paper, \citet{Rowe2014} confirmed the inner and outer planets, PH3~b and d, via ``validation by multiplicity'' based on the statistical assessment that multiple transiting signals are unlikely to be false positives.  They were given the names Kepler-289 b and c.  See Table~\ref{tab:names} for cross-matching the names of the planets.   Furthermore, PH3 d's TTVs were detected at a significant level in \citet{Mazeh2013}, but they were only able to fit the signal to a parabola rather than a full sine curve.  As such, they do not measure the TTV period or its amplitude.  PH3~c's TTVs were undetected because \citet{Mazeh2013} only searched through KOIs. This system has period ratios of 1.91 (both the outer/middle and middle/inner period ratios), which is somewhat far from the 1:2:4 mean motion resonance (MMR).  

\begin{deluxetable}{lccc}
\tablewidth{0pt}
\tablecaption{Stellar and planetary designations. \label{tab:names}}
\tablehead{
\colhead{Star} &
\colhead{Inner planet} &
\colhead{Middle planet} &
\colhead{Outer planet} }
\startdata

PH3          & PH3~b         & PH3~c        & PH3~d         \\ 
Kepler-289   & Kepler-289~b  & \nodata      & Kepler-289~c  \\ 
KOI-1353     & KOI-1353.02   & \nodata$^*$  & KOI-1353.01   \\ 

\enddata

\tablecomments{$^*$ Identified as a TCE with a period five times the true 66 day period.}

\end{deluxetable}  

\subsection{Transit Timing Variations and Mean Motion Resonances}
\label{sec:ttv}

Since the transit technique provides a planet's radius, a mass measurement can both confirm the planetary nature of the candidate as well as measure the planetary density.  In some systems with multiple planets, TTVs can determine or provide constraints on the planet masses \citep{Miralda2002,Agol2005,Holman2005}.   TTVs of transiting  planets can even be used to indicate the presence of non-transiting planets \citep{Ballard2011,Nesvorny2012}, co-orbital (Trojan) planets \citep{Ford2007}, or exomoons \citep{Kipping2009}.  TTVs were first used to confirm a pair of planet candidates orbiting Kepler-9 \citep{Holman2010}. Since then, numerous other \textit{Kepler} candidates have been confirmed using TTVs \citep[e.g.,][]{Lissauer2011,Steffen2012,Xie2014}.

One potential disadvantage of measuring planet masses with TTVs is that, in some circumstances, the masses become degenerate with the eccentricity \citep{Lithwick2012}.  This occurs when: 1) the planet system is close to, but not within, a first-order mean motion resonance; and 2) the measurement errors are significant enough to only detect the sinusoidal TTV caused by the nearby resonance.   In this case, the amplitude of the TTV depends both on the planet masses and on the free eccentricity, $z_{\rm{free}}$, which is the component of eccentricity that is not driven  by the resonant forcing.  If the $z_{\rm{free}}$ is known, the mass is also known.  \citet{Lithwick2012} examines the degeneracy closely for systems near MMR. In a $z_{\rm{free}}=0$ system, the TTVs of pairs of interacting planets will be anti-correlated.  If the two sets of TTVs are not anti-correlated, this implies that significant free eccentricities exist.  However, the absence of a phase shift does not necessarily mean that free eccentricities are zero \citep{Lithwick2012}.  This could instead be an unlikely moment in time when free eccentricities do exist, but their phases cancel out.  If $z_{\rm{free}}$ is small, then one can calculate an upper limit on the mass by assuming $z_{\rm{free}}=0$ \citep{Lithwick2012}.  Determining $z_{\rm{free}}$ allows one to calculate the true mass. 

The mass-$z_{\rm{free}}$ degeneracy can also be broken with more precise data.  An $O-C$ TTV signal for a pair of planets near MMR results in a periodic signal (sinusoidal for cases of low $z_{\rm{free}}$).  This period is equal to:

\begin{center}	
\begin{equation}
\label{eq:ttv}
P_{\rm{TTV}}=\frac{1}{j/P_j - (j-1)/P_{j-1}}
\end{equation}
\end{center}

\noindent where the outer and inner planets are in a near $j:j-1$ resonance, respectively, and $j$ is a positive integer \citep{Agol2005,Lithwick2012}.  This period is called the super-period or TTV period.  After one fits the transit midpoints with a linear period and the TTV period, a combination of high quality data and large amplitude TTVs can show the proportionally smaller residual chopping signal \citep{Agol2005,Fabrycky2012b}.  This chopping signal is best seen for the inner planet of a near-resonant system and is not seen as strongly in the outer planet, as the position of the inner planet at the time of the outer planet's transit changes very slowly when near a $j:j-1$ MMR.  The chopping signal scales with the mass of the outer planet \citep{Holman2010} and with the eccentricity \citep{Nesvorny2014}.  This additional constraint allows one to break the mass-$z_{\rm{free}}$ degeneracy to calculate the mass of each planet.

\section{Orbital and Stellar Characterization}

\subsection{Orbital fit}

\begin{figure*}[tbp]
	\centering
		\includegraphics[width=1.00\textwidth]{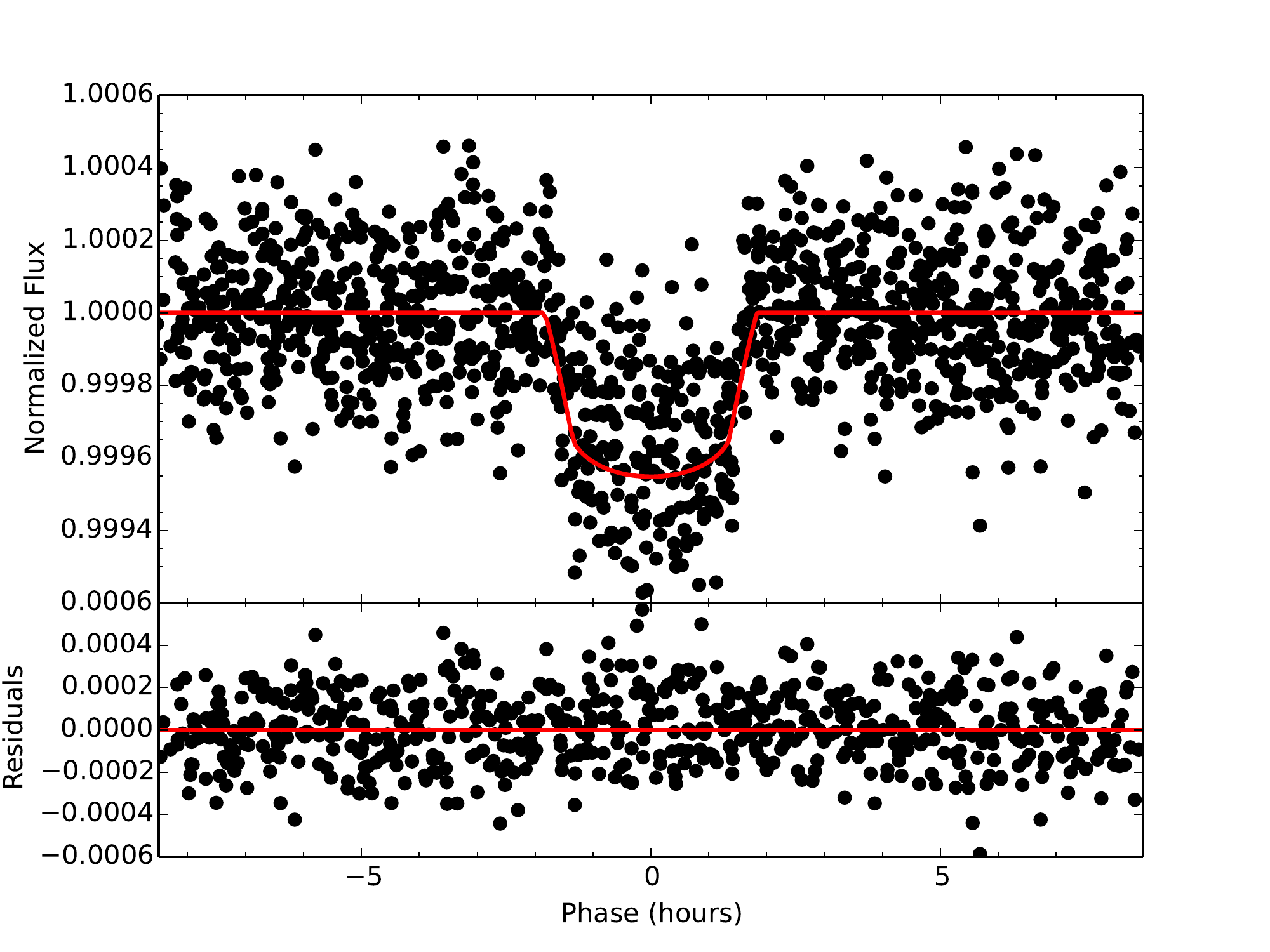}
	\caption{The top panel shows the inner planet's (PH3~b's) phase-folded transit light curve (black data points) with the model overplotted in red.  The residuals are shown in the bottom panel.}
	\label{fig:intransit}
\end{figure*}

\begin{figure*}[tbp]
	\centering
		\includegraphics[width=1.00\textwidth]{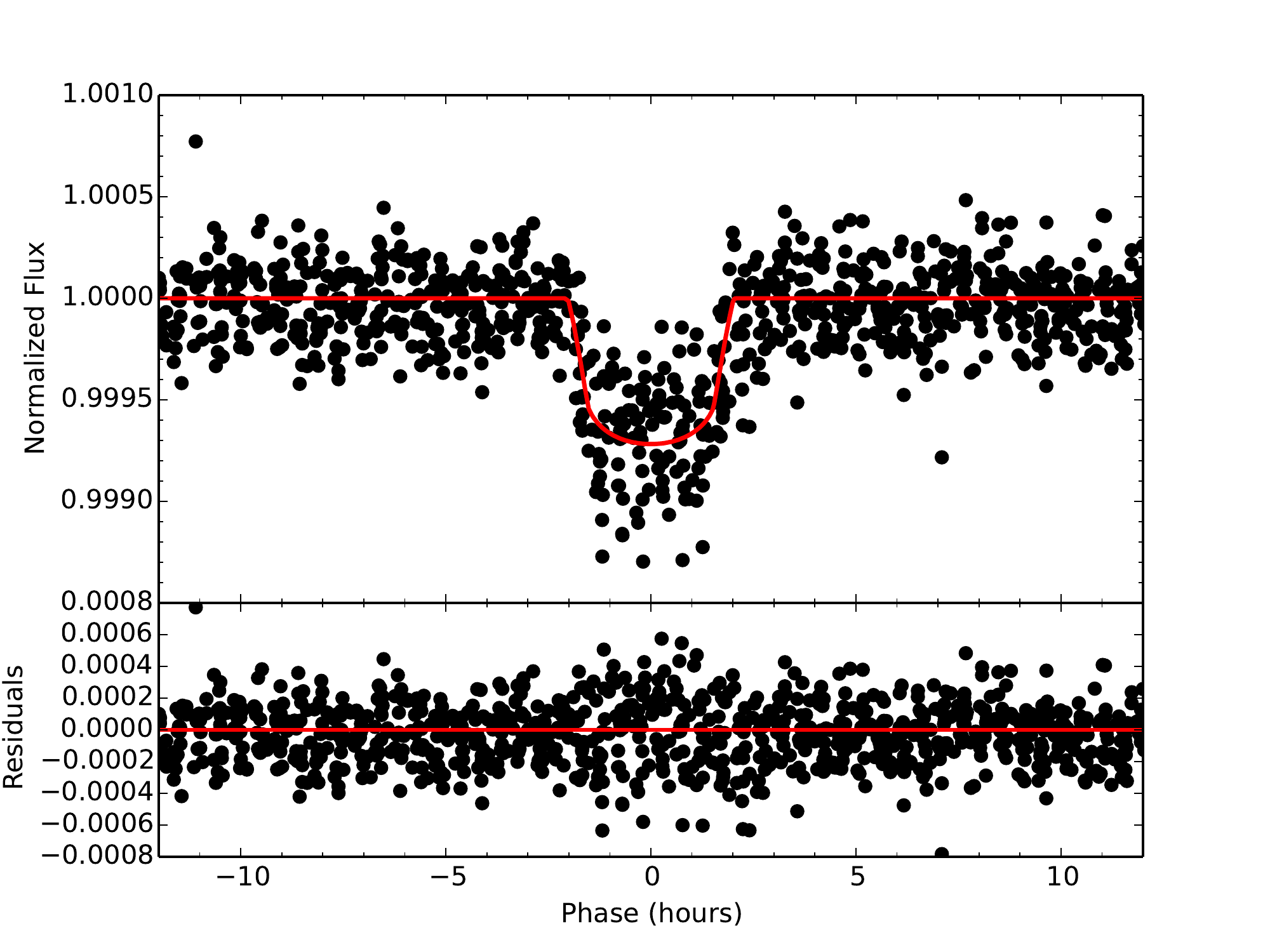}
	\caption{The top panel shows the middle planet's (PH3~c's) phase-folded transit light curve (black data points) with the model overplotted in red.  The residuals are shown in the bottom panel.  The extra scatter in the residuals are due to starspot anomalies.}
	\label{fig:midtransit}
\end{figure*}

\begin{figure*}[tbp]
	\centering
		\includegraphics[width=1.00\textwidth]{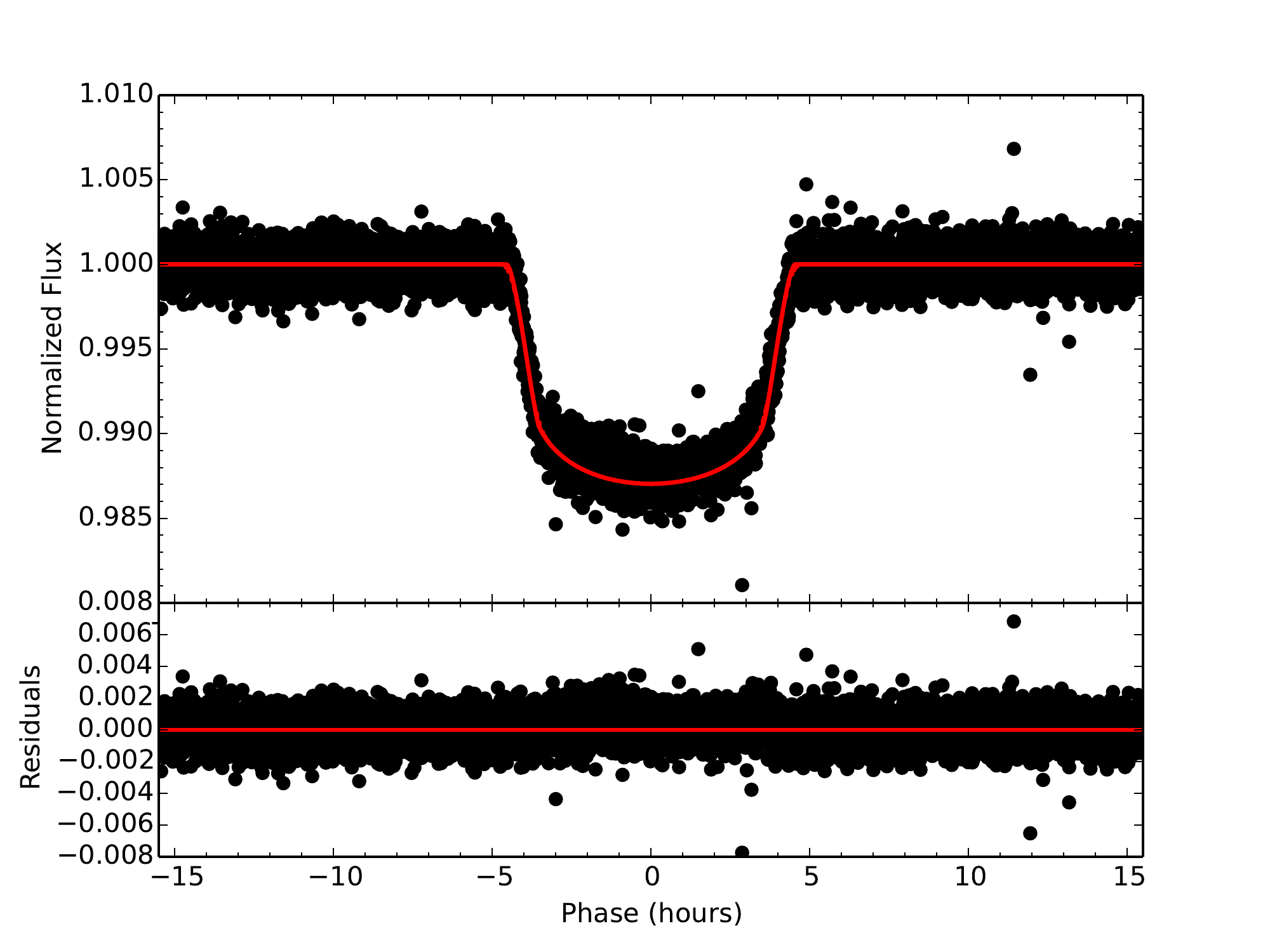}
	\caption{The top panel shows the outer planet's (PH3~d's) phase-folded transit light curve (black data points) with the model overplotted in red.  The residuals are shown in the bottom panel.  There exists a slight residual due to starspots, primarily small-scale starspot crossings. Small spots in the ninth transit produce the majority of this effect, and due to its high density of short cadence data points, the residual is more readily noticeable.}
	\label{fig:outtransit}
\end{figure*}

Using quarters 1-16 of the \textit{Kepler} data, we extracted and flattened each transit using the IDL \texttt{AutoKep} program \citep{Gazak2012}.  For the high signal-to-noise transits of the outer planet, we used short cadence data where available.  We then used a new, modified version of the IDL program \texttt{TAP} \citep{Carter2009,Gazak2012,Eastman2013} to fit for the orbital parameters of each planet:  impact parameter ($b$), duration ($T$), the ratio of planet radius to stellar radius ($R_{\rm{p}}/R_*$), the midtransit times, linear limb darkening, quadratic limb darkening \citep{Kipping2013}, and white and red noise.  The ratio of semi-major axis to the radius of the star ($a/R_*$) and the inclination ($i$) can be derived from these parameters.  For the purpose of transit fitting, circular orbits were assumed, which we found to be a good assumption (see Section~\ref{sec:agol}).  See Figures~\ref{fig:intransit}, \ref{fig:midtransit}, and \ref{fig:outtransit} for the transit fits of PH3~b, c, and d, respectively.

The linear and quadratic limb darkening is very poorly constrained for PH3~b and c when analyzing each planet individually. Therefore, we used the posterior distribution for the linear and quadratic limb darkening from the high signal-to-noise transits (depth $\approx12500$~ppm) of PH3~d's fit as a prior for the inner and middle transit fits.  Starspot anomalies are seen in a few transits of PH3~d.  We masked out obvious spots, but we did not account for smaller scale starspot anomalies. This causes a slight residual in the light curve of PH3~d, primarily due to the ninth transit (see Figure~\ref{fig:outtransit}). For PH3~b, \texttt{TAP} was unable to simultaneously fit for both midtransit times and the orbital properties.  Therefore, we fit the phase-folded model first for the orbital parameters.  Then, holding the orbital parameters fixed, we fit for the midtransit times. 

The parameter $a/R_*$ from the \texttt{TAP} fit is poorly constrained for PH3~b and PH3~c, and their best-fit values are inconsistent with PH3~d's best fit.  For example, PH3~c's $a/R_*$ is greater than PH3~d's, but it is on an interior orbit (see Table~\ref{tab:planets}).  Therefore, we revise the $a/R_*$ for each planet using Kepler's Third Law with the planet's period and the stellar parameters $M_*$ and $R_*$ (derived from stellar density; see Section~\ref{sec:stellarfit}) to get $(a/R_*)_{\rm{rev}}$.  The two values agree within errors, the unrevised $a/R_*$ approximately one $\sigma$ higher than $(a/R_*)_{\rm{rev}}$.  We use the revised $a/R_*$ to calculate the planet's incident flux ($S$) and semi-major axis ($a$).  The best fit transit light curves for PH3~b, c, and d are shown in Figures~\ref{fig:intransit}, \ref{fig:midtransit}, and \ref{fig:outtransit}, respectively.  The orbital and planetary properties are shown in Table~\ref{tab:planets}.

\begin{deluxetable*}{lccc}
\tablewidth{0pt}
\tablecaption{Orbital and planetary parameters of PH3 b, c, and d. \label{tab:planets}}
\tablehead{
\colhead{} &
\colhead{PH3~b} &
\colhead{PH3~c} &
\colhead{PH3~d} }

\startdata

$P$ (days)                          & $34.5450\pm0.0005$            & $66.0634\pm0.0114$            & $125.8518\pm0.0076$              \\ 
$T_0$ (JD$-$2454000)                & $965.6404\pm0.0040$           & $975.6436\pm0.0068$           & $1069.6528\pm0.0077$             \\ 
$M_{\rm{P}}$ ($M_\oplus$)           & $7.3\pm6.8$                   & $4.0\pm0.9$                   & $132\pm17$                       \\ 
$R_{\rm{P}}$ ($R_\oplus$)           & $2.15\pm0.10$                 & $2.68\pm0.17$                 & $11.59\pm0.19$                   \\ 
$\rho$ ($g/cm^3$)                   & $4.1\pm3.9$                   & $1.2\pm0.3$                   & $0.47\pm0.06$                    \\ 
$a_{\rm{rev}}$ ($AU$)               & $0.21\pm0.01$                 & $0.33\pm0.02$                 & $0.51\pm0.03$                    \\
$S$ ($S_\oplus$)                    & $24.8\pm4.4$                  & $10.7\pm1.8$                  & $4.4\pm0.8$                      \\
$t_{\rm{dur}}$ (hours)              & $3.178^{+0.055}_{-0.053}$     & $3.557^{+0.072}_{-0.065}$     & $8.067^{+0.026}_{-0.024}$        \\ 
$a/R_*$                             & $71.1^{+10}_{-20}$            & $117.8^{+21}_{-42}$           & $109.5\pm1.2$                    \\ 
$(a/R_*)_{\rm{rev}}$                & $45.9\pm0.5$                  & $70.7\pm0.7$                  & $108.6\pm1.1$                    \\
$b$                                 & $0.04^{+0.66}_{-0.68}$        & $0.05^{+0.70}_{-0.76}$        & $0.394^{+0.026}_{-0.029}$        \\ 
$i$ (deg)                           & $89.59^{+0.30}_{-0.48}$       & $89.73^{+0.20}_{-0.38}$       & $89.794^{+0.017}_{-0.016}$       \\ 
$R_{\rm{P}}/R_*$                    & $0.0197^{+0.0011}_{-0.0006}$  & $0.0246^{+0.0022}_{-0.0009}$  & $0.10620^{+0.00049}_{-0.00050}$  \\ 
$M_{\rm{P}}/M_*$ ($\times10^{-5}$)  & $2.0\pm1.9$                   & $1.1\pm0.2$                   & $36.43\pm4.66$                   \\
$e\cos(\omega)$                     & $-0.0215\pm0.0255$            & $-0.0035\pm0.0022$            & $0.0032\pm0.0066$                \\
$e\sin(\omega)$                     & $-0.0113\pm0.0239$            & $-0.0108\pm0.0122$            & $0.0033\pm0.0086$                \\
 
\enddata

\tablecomments{Best-fit parameters for the orbital and planetary properties.  The period is the mean period given over the length of observations.  The best fit $a/R_*$ from \texttt{TAP} is poorly constrained for PH3~b and PH3~c and are obviously inconsistent with PH3~d.  Therefore, we revise it using the Newton's version of Kepler's Third Law with $M_*$, $R_*$, and $P$ to get $(a/R_*)_{\rm{rev}}$.}

\end{deluxetable*}

\subsection{Stellar fit}
\label{sec:stellarfit}

We derived the stellar density, $\rho_*$, from the Markov chain orbital fit analysis of the outer planet, converting the observed transit characteristics into the density of the star at each link in the chain to derive the posterior distribution of the stellar density \citep{Seager2003}.  In computing the density, we calculate the sky velocity of the planet during transit; this in turn depends on the eccentricity, $e$, and longitude of periastron, $\omega$, or equivalently the eccentricity vector, $e = (e\cos{\omega},e\sin{\omega})$. Each component of the eccentricity vector of the planet was constrained by the standard deviation of its posterior distribution derived from the transit-timing analysis, which was found to have a negligible effect on stellar density (see Section \ref{sec:agol}).  This resulted in a stellar density of $\rho_* = 1.577 \pm 0.066$ g cm$^{-3}$.

We obtained a spectrum of KIC~7303287 with the HIRES instrument on the Keck I telescope \citep{Vogt1994} and performed Spectroscopy Made Easy (SME) on the data \citep{Valenti1996,Valenti2005}.  To further constrain the stellar mass ($M_*$), radius ($R_*$), and age ($t_*$), we fit the Padova PARSEC (v1.1) isochrone models \citep{Bressan2012} to the observed properties of the PH3 host star: 1) the effective temperature ($T_{\rm{eff}}$) and metallicity ($\rm{[Fe/H]}$) derived from the SME analysis; 2) the stellar density, $\rho_*$ (rather than $\log{g}$, as $\log{g}$ is poorly constrained and weakly dependent on the mass of the star on the main sequence);  3) the spectral energy distribution derived from measured stellar magnitudes in three available databases:  SDSS g, r, i, and z \citep{Abazajian2009}, 2MASS J, H, and Ks \cite{Cutri2003}, and WISE W1 and W2 \citep{Cutri2012}. For each band we used the reported uncertainty, $\sigma_i$, but also accounted for systematic uncertainty in the calibration of the flux by adding in quadrature a magnitude uncertainty, $\sigma_0$, that we allowed to float, and multiplied the likelihood by $\Pi_i(\sigma_i^2+\sigma_0^2)^{-1/2}$ \citep{Ford2006}.  We assumed a uniform prior on the extinction, $A_V$, and chose an $R_V=3.1$ Milky Way extinction law.  We also allowed the distance, $D$, and $t_*$ to vary.

There are six stellar model parameters: ($M_*$, $\rm{[Fe/H]}$, $t_*$, $D$, $A_V$, $\sigma_0$). We computed a Markov chain using the affine-invariant population Markov chain method \citep{Goodman2010,Foreman-Mackey2013} with eighteen chains (three times the number of free parameters). At each point in the chain, the likelihood was computed from the Gaussian probability of the agreement with the nine observed magnitudes, the stellar density, the effective temperature, and the metallicity.  To compute these properties, the stellar isochrones were linearly (or log-linearly) interpolated from a grid of stellar models. The stellar isochrones assumed scaled solar abundances ($\rm{[}\alpha\rm{/Fe]}=0$) and sampled metallicity in intervals of 0.1 dex, while we sampled age in intervals of 0.05 dex and mass in intervals that vary with age and metallicity.  Table~\ref{tab:star} gives the results of this analysis, which show that this is a young, solar-type star.  These stellar parameters are consistent with those derived by \citet{Rowe2014} and are moderately consistent with \citet{Huber2014}, which are also shown for comparison.   The small uncertainties on the stellar parameters are due to the precise determination of the density from the light curve and transit timing analysis and from the precise temperature and metallicity measurement from SME.  However,  these uncertainties do not account for possible systematic errors in the analysis.  Hence, we re-ran our isochrone analysis using the Dartmouth isochrones \citep{Dotter2008} and found results that were consistent within $1\sigma$ in the mean and yielded standard deviations of similar magnitude.  Also, see \citet{Torres2012} for a discussion of the systematic error and biases of SME and a comparison to stellar parameter classification (SPC) technique and MOOG.

\begin{figure*}[tbp]
	\centering
		\includegraphics[width=1.00\textwidth]{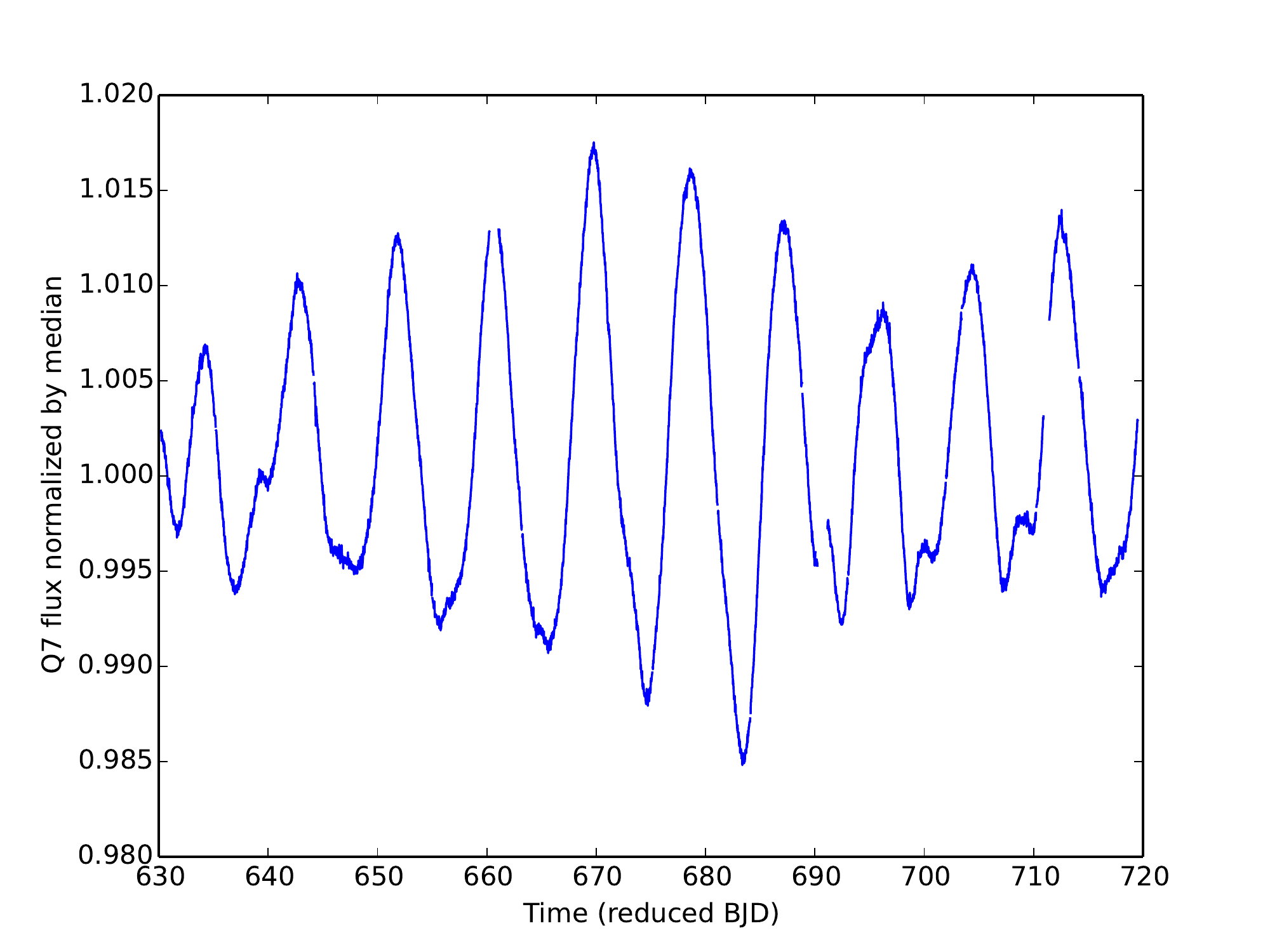}
	\caption{The normalized light curve of PH3 during quarter 7 as a function of the reduced barycentric Julian day (BJD).  The strong variability is caused by starspots and can be used to calculate its rotation period.}
	\label{fig:star}
\end{figure*}

PH3 shows strong activity clearly attributable to starspots (see Figure~\ref{fig:star}).  The long, deep duration of PH3~d shows several strong starspot anomalies, with the starspots primarily appearing at phases of about 1.2 hours before and 3.5 hours after the transit midpoint (see Figure~\ref{fig:outtransit}).  This may allow for a future investigation of its spin-orbit obliquity.  Starspots may also cause small, minute-scale TTVs themselves \citep{Mazeh2013}.  However, the orbital period of the outer planet is long compared to the rotational period, making this investigation difficult and outside the scope of our study.  The starspots allow for a determination of the rotational period, $t_{*,\rm{rot}}=8.648\pm0.0009$ days \citep{McQuillan2013}.   Gyrochronology allows one to calculate the age of the star ($t_{*,\rm{gyro}}$) using corrected $B-V$ colors and the stellar rotation period \citep{Barnes2007}.  Using the rotational period, the $B$ and $V$ magnitudes from the MAST\footnote{http://archive.stsci.edu/index.html} catalog ($B=14.816$, $V=14.255$), the earlier derived $A_V=0.06$, and the Milky Way extinction law ($R_V=3.1$), we find $t_{*,\rm{gyro}}=1.0\pm0.3$ Gyr, which is somewhat larger but consistent with the age derived from the Padova isochrones.

\begin{deluxetable*}{lccc}
\tablewidth{0pt}
\tablecaption{Stellar parameters of PH3 (KIC 7303287). \label{tab:star}}
\tablehead{
\colhead{KIC 7303287}        &
\colhead{This paper}         &
\colhead{\citet{Huber2014}}  &
\colhead{\citet{Rowe2014}}   \\ }

\startdata

$M_*$ ($M_\odot$)        & $1.08\pm0.02$    & $1.16^{+0.31}_{-0.17}$   & $1.04^{\dagger}$  \\ 
$R_*$ ($R_\odot$)        & $1.00\pm0.02$    & $1.60^{+0.83}_{-0.48}$   & $1.16\pm0.22$     \\ 
$T_{\rm{eff}}$ (K)       & $5990\pm38$      & $6279^{+171}_{-215}$     & $5930\pm107$      \\ 
$\log{g}$ (cgs)          & $4.47\pm0.01$    & $4.10^{+0.25}_{-0.27}$   & $4.33\pm0.15$     \\ 
$\rho_*$ ($g/cm^3$)      & $1.58\pm0.07$    & $0.40^{+0.61}_{-0.26}$   & $0.94\pm0.42$     \\ 
$\rm{[Fe/H]}$            & $0.05\pm0.04$    & $-0.08^{+0.24}_{-0.30}$  & $-0.06\pm0.10$    \\ 
$L_*$ ($L_\odot$)        & $1.15\pm0.06$    & $3.58^{\dagger}$         & $1.50^{\dagger}$  \\ 
$t_*$ (Gyr)              & $0.65\pm0.44$    & \nodata                  & \nodata           \\
$t_{*,\rm{gyro}}$ (Gyr)  & $1.0\pm0.3$      & \nodata                  & \nodata           \\
$D$ (pc)                 & $700\pm14$       & \nodata                  & \nodata           \\
$A_V$                    & $0.06\pm0.01$    & \nodata                  & \nodata           \\
 
\enddata

\tablecomments{For comparison, we also show the stellar parameters from \citet{Huber2014} and \citet{Rowe2014}.  We note that \citet{Rowe2014} used a different Keck HIRES spectrum than us as their observational input for stellar parameters. \\
$^{\dagger}$ Parameter was not explicitly given, so calculated from other parameters as appropriate for the sake of comparison.  Due to the unknown posterior distributions of these values, we do not propagate its error. }

\end{deluxetable*}

\section{Transit Timing Mass Determinations}
\label{sec:agol}

Interactions between transiting planets can result in TTVs, which can provide constraints on the orbital parameters and masses of the transiting planets \citep{Agol2005,Holman2005}.  We modeled the times of transit with TTVFast, a newly written code for computing transit times of multi-planet transiting systems \citep{Deck2014}.  A symplectic integrator was used to compute the positions of the planets as a function of time \citep{Wisdom1991,Wisdom2006}, while Keplerian interpolation was used to find the times of transit \citep{Nesvorny2013}.   The three planets were assumed to be on plane-parallel orbits,  and the likelihood was computed using a $\chi^2$ fit to the times of transit, with the error bars of each planet inflated by a constant factor, $f_i$, which was added to the model. For each planet, five parameters were used to describe the orbit: the time of the first transit, $t_i$, the initial orbital period, $P_i$, the eccentricity vector, ($e_i\cos{\omega_i}$,$e_i\sin{\omega_i}$), and the mass ratio of the planet to the star, $\mu_i$.  With $i=1, 2, 3$, there are a total of 18 model parameters.  A prior on the likelihood of $f_i^{-N_i/2}$, with $N_i$ being the number of transits of planet $i$, was added to penalize large values of $f_i$;  this resulted in a median $\chi^2_{\rm{red}}$ for each planet of order unity \citep{Ford2006}.  A prior of $1/e_i$ was added for each planet to prevent small eccentricities from being disfavored \citep{Ford2006}.  One transit for each PH3~b and PH3~c were excluded as outliers with their contribution to $\chi^2\gg1.0$; their midtransit times were approximately $2456209$ and $2455900$ JD, respectively.

\begin{figure*}[tbp]
	\centering
		\includegraphics[width=0.90\textwidth]{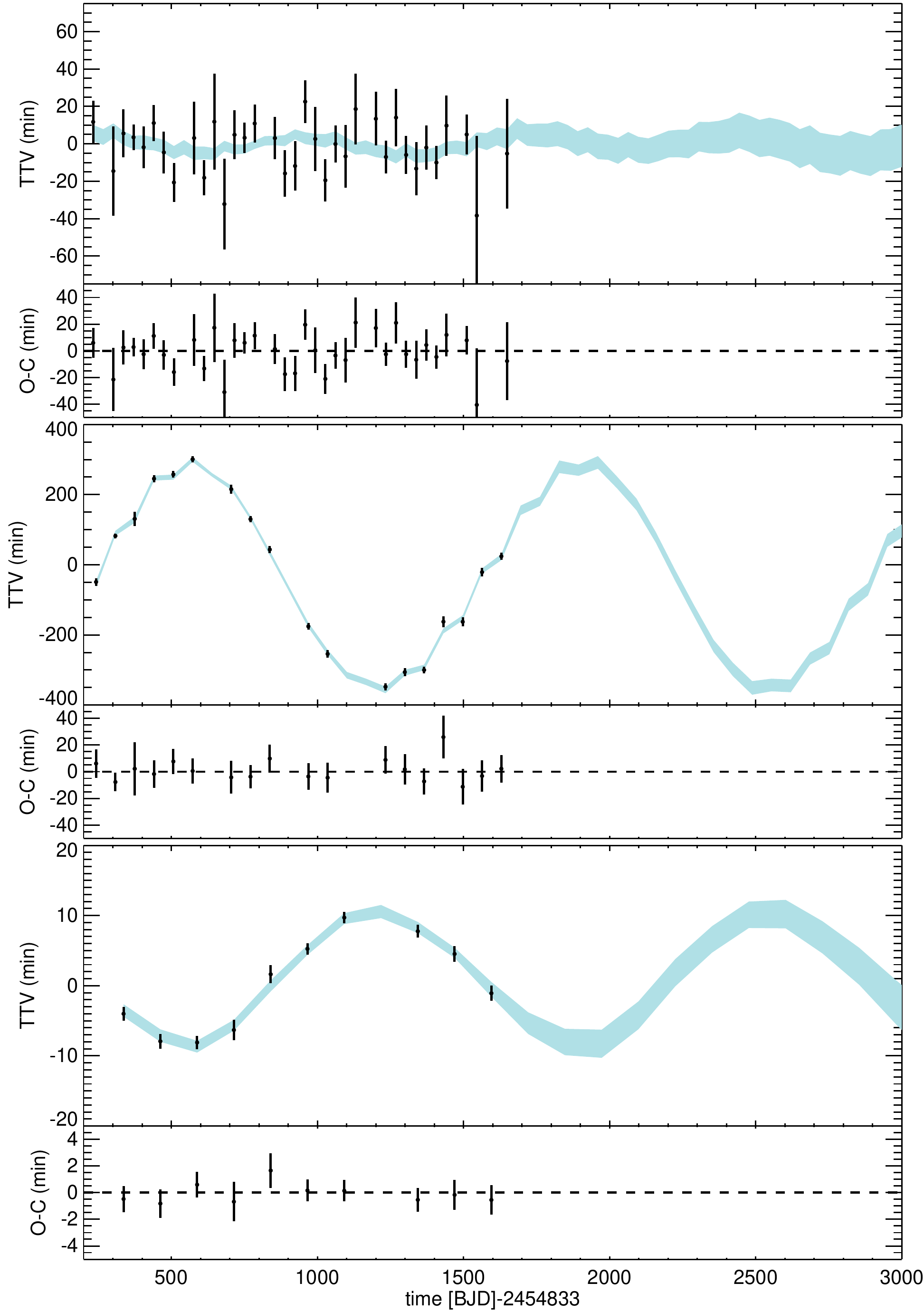}
	\caption{The TTVs of each planet (top: PH3~b, middle: PH3~c, bottom: PH3~d) and the residuals from the fit.  Notice the sinusoidal chopping signal PH3~c.  This breaks the mass-eccentricity degeneracy allowing for a unique determination of mass. The blue shaded region is the one sigma best-fit region.}
	\label{fig:ttvs}
\end{figure*}

\begin{figure*}[tbp]
	\centering
		\includegraphics[width=0.90\textwidth]{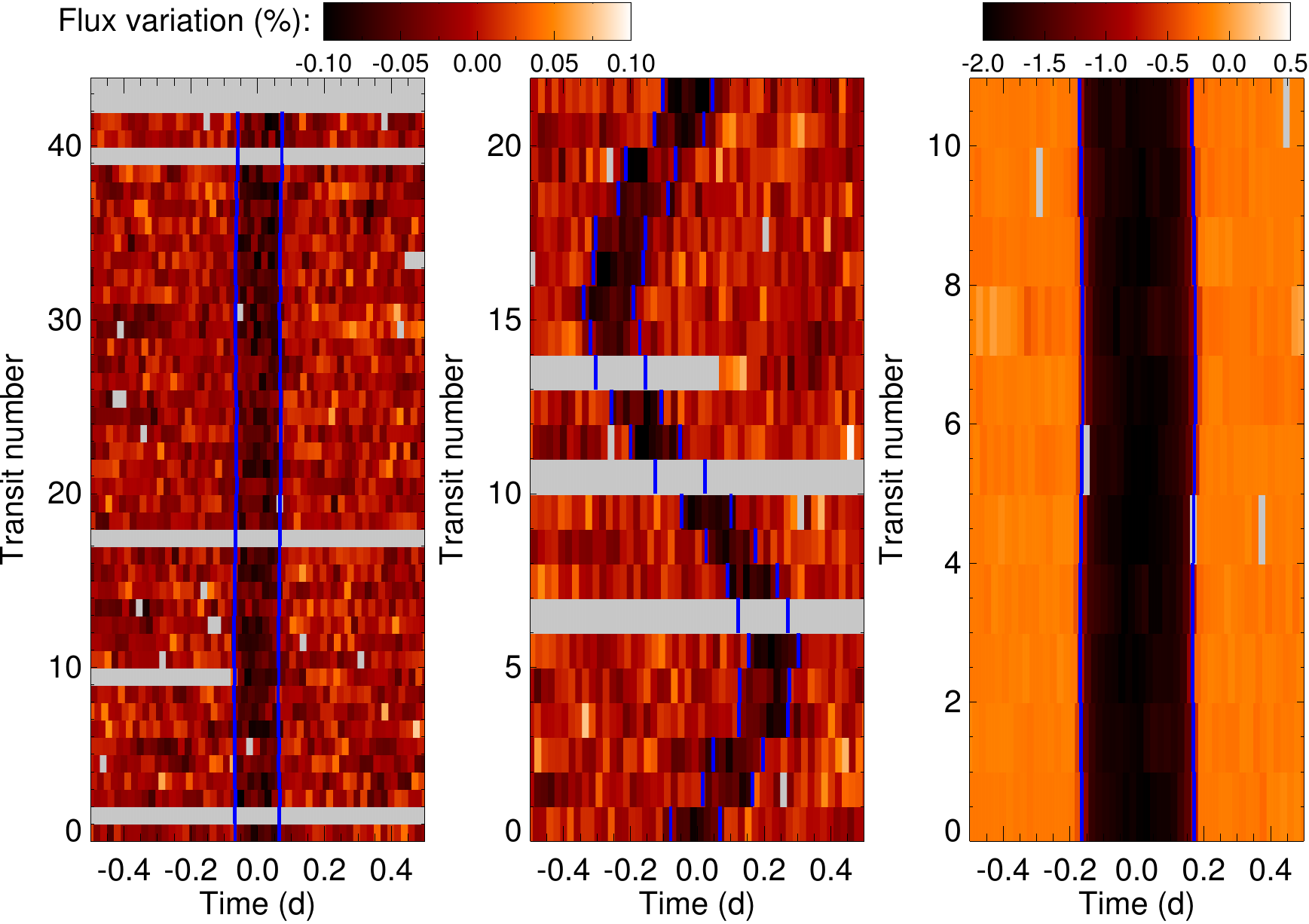}
	\caption{The TTVs of each planet (left: PH3~b, middle: PH3~c, right: PH3~d) in ``lava plot'' (or ``river plot'') format. The colored pixels represent the relative flux, increasing from black to red to white.  Each row of the plot shows a segment of the light curve around each transit, starting from the first transit on the bottom to the top transit at the top.  The ingress and egress of each transit is marked with blue lines, and data gaps are represented by grey bars.  Transit timing variations manifest themselves as a dark path winding back and forth in the lava and are readily seen in PH3~c.}
	\label{fig:lava}
\end{figure*}

An affine-invariant ensemble  MCMC code was used to compute the posterior distribution on the parameters \citep{Goodman2010,Foreman-Mackey2013}.  An ensemble of 31 chains was run for $10^6$ generations, reaching a high degree of convergence; the first 60,000 generations were discarded when computing the posterior parameters.  Figure~\ref{fig:ttvs} shows the resulting transit-timing variations of the three planets with the 1$\sigma$ confidence intervals computed from the chain;  the plotted errors are inflated by $f=(1.73, 1.79, 1.35)$ for the inner to outer planets, respectively.  There is evidence of starspot crossings in the transit light curves, which we attempted to address by masking these regions in the light curve modeling.  However, this approach cannot take into account the effect of smaller spots crossed by the planets during a transit that lead to variation in the depth of transit that is commensurate with the photometric uncertainties.  Such spot crossings do not affect the out-of-transit data, and hence are not accounted for in the red-noise model used by TAP.  Spot crossings cause the transit shape to be asymmetric, and hence can skew the fit for the mid-transit time if not accounted for.  Hence, we include $f_i$ to account for additional systematic uncertainties on the errors in the times of transit that are not accounted for in our light curve fitting. Figure~\ref{fig:lava} shows a ``lava plot'' (or ``river plot''), demonstrating the wavy nature of TTVs.  The results show that the phase of the 1325 day TTV super period is not \textit{exactly} anti-correlated, requiring a small free eccentricity in the system, but this has a neglible contribution to $\rho_*$.  Table~\ref{tab:planets} gives the best-fit parameters for the planets' orbital parameters and masses.  All three planet masses are well within the planetary regime, confirming them as true planets.  

The outer two planets in KOI-1353 are dynamically interacting, and this effect is large since they are close to a $2:1$ period ratio.  Their TTVs appear as anti-correlated sinusoidal variations with a period of $P_{TTV} = 1/(2/P_3-1/P_2) = 1325\pm 5$ days, with $P_2=66.0634\pm 0.0114$ days, and $P_3=125.8518\pm 0.0076$ days (see Equation~\ref{eq:ttv} and Figure~\ref{fig:ttvs}).  These two planets are near commensurate, but not in resonance.  The observed chopping signal breaks the mass-eccentricity degeneracy for this pair of planets.  Since the middle planet has the largest amplitude TTV, the mass of the outer planet is constrained more precisely than the mass of the middle planet. 

\section{Long-term Stability}

We tested the long-term stability of the system by numerically integrating a random set of 1000 initial conditions which produced TTVs consistent with the \textit{Kepler} data. These 1000 initial conditions and corresponding planetary masses were drawn from the posterior distribution resulting from fitting the \textit{Kepler} data using Markov Chain Monte Carlo and hence form a statistically representative set of solutions.  We integrated these solutions for $10^8$ orbits of the inner planet, or $\sim 10^7$ years, using a Wisdom-Holman mapping \citep{Wisdom1991} in combination with a third-order symplectic corrector \citep{Wisdom2006}. We employed a time step of one day which resulted in a maximum fractional energy error of $\sim \times10^{-10}$. 

We then determined the maximum full oscillation amplitude for the semi-major axis and eccentricity of each of the planets. Larger variations in semi-major axis compared to the average value are a sign of instability, whereas large eccentricity oscillations may result simply from large free eccentricities and, in this case, do not indicate unstable behavior. During the $10^8$ orbits of the inner planet, we only found deviations of less than a percent in the semi-major axes. The median eccentricity oscillations of the inner and outermost planet during these integrations were on the order of $30\%$ of the mean value, but the middle planet had large eccentricity oscillations of full amplitude $\sim150\%$ the mean value. Out of all the initial conditions, the maximum eccentricity oscillations were on the order of $200\%$ of the mean value.  These results are summarized in Table \ref{tab:kat}.

\begin{deluxetable*}{lccc}
\tablewidth{0pt}
\tablecaption{Results for semi-major axis and eccentricity oscillations about the mean ($\overline{a}$ and $\overline{e}$, respectively) after $10^8$ orbits of PH3~b \label{tab:kat}}
\tablehead{
\colhead{Planet}                          &
\colhead{Max [$(\Delta a)/\bar{a}] (\%)$}     &
\colhead{Median [$(\Delta e)/\bar{e}] (\%)$}  &
\colhead{Max [$(\Delta e)/\bar{e}] (\%)$}     \\ }

\startdata

PH3~b  & 0.0443259  & 34.0028  & 219.367  \\ 
PH3~c  & 0.302436   & 143.627  & 212.864  \\ 
PH3~d  & 0.0144348  & 27.6211  & 202.432  \\ 

\enddata

\end{deluxetable*}

The integrations indicated that the orbits were quiescent. We followed a subset of 100 of them for 10 times longer, for $10^9$ orbits of the inner planet or $10^8$ years. We find no significant deviations from the oscillation amplitudes reported in Table \ref{tab:kat}.

Although there is no analytic criterion for the stability of three planet systems, one can gain some insight by studying each pair of planets individually. Both the inner pair and outer pair satisfy the two-planet Hill criterion for stability \citep[e.g.,][]{Gladman1993} as well as the heuristic three-planet stability criterion used by \citep{Fabrycky2012a} for other \textit{Kepler} systems. Furthermore, neither pair is near a low-order mean motion resonance.  Although integrations of a length shorter than the age of the system cannot prove the system's stability, our tests strongly suggest that these orbits are long-lived.

\section{Laplace resonance}

With such low eccentricities, masses, and period ratios of 1.91, neither pair of planets is close enough to the 2:1 mean motion resonance for it to substantially affect the dynamics of the system. It is interesting to also establish how close the system is to a Laplace (or three-body) resonance. Note, however, that typically, three-body resonances are only important if one or both of the pairs is itself very close to a two-body resonance. 

The equation of motion for the three-body resonance angle of the form $\phi = p \lambda_1 -(p+q) \lambda_2 +q \lambda_3$, where $p$ and $q$ are integers and $\lambda_i$ is the mean longitude of the planet $i$, resembles at lowest order the angle of a simple pendulum \citep{Aksnes1988,Quillen2011}. The Hamiltonian for this is written as:

\begin{equation}
H = \frac{1}{2} A_{p,q} p_\phi^2 +\epsilon_{p,q} \cos{\phi}
\end{equation}

\noindent where the coefficients $A_{p,q}$ and $\epsilon_{p,q}$ depend on which $(p,q)$ resonance is being considered. Therefore, one can approximate the full width of the resonance region (where the angle $\phi$ oscillates) simply as $\Delta p_\phi \sim 4 \sqrt{\epsilon_{p,q}/A_{p,q}}$. This can then be converted into a width in terms of the semi-major axis of one of the planets. 

We follow the work of \citet{Quillen2011}, which focuses on the three-body resonances of circular orbits not near any two-body resonances and which shows that for equal mass planets the width of the resonance in terms of the semi-major axis of the inner planet $a_1$ roughly scales as:

\begin{equation}
\label{eq:a1}
\frac{\Delta a_1}{a_1} \sim \frac{m}{m_\star} \frac{|\log{\delta}|}{\delta} \exp{[-(p+q)\frac{\delta}{2}]}
\end{equation}

\noindent with $m$ the mass of one of the planets and for semi-major axis ratios of order $\alpha \sim 1-\delta$. In other words, for close pairs of planets the width of the resonance depends linearly on the mass of the planet, relative to the mass of the star, and is significantly larger for closer pairs of planets (smaller $\delta$). Moreover, unless $\delta$ is very small, the widths of the resonances with large values of $p$, $q$ will be exponentially small. This already suggests that the Laplace resonance will not be important for PH3.

We follow the exact formulas given in \citet{Quillen2011}, which do not assume equal mass planets, for $\epsilon_{p,q}$ and $A_{p,q}$ to determine how far away the system PH3 is from each $(p,q)$ three-body resonance. Given the orbital periods of the outer two planets and the masses of all of them, we determine the semi-major axis $a_{1,r}$ that the innermost planet would need such that the system would be in exact resonance (satisfying $\dot{\phi} = 0$), and we also determine the width of the resonance in terms of the semi-major axis of the innermost planet $\Delta a_1$.  We then computed the number $(a_1 -a_{1,r})/\Delta a_1$, the dimensionless number of how many resonance widths in semi-major axis of the inner planet the system was away from exact resonance, for each $(p,q)$ pair, with $0<p$, $q<30$. 


As expected from Equation~\ref{eq:a1}, the resonances with larger values of $(p,q)$ are entirely negligible. Their widths are too small for the system to be near any of them. The PH3 system is closest to the three body resonances of the form $(p,q) = i(1, 2)$, with $i$ being an integer greater or equal to unity. However, the system is already $\sim 60$ widths away even from the $(1, 2)$ resonance and is orders of magnitude further from the majority of the resonances looked at.  From this, we conclude that three-body resonances in this regime (not near a two-body resonance and with circular orbits) are unimportant for this system.

\section{Planet Composition}

To constrain the masses of the interior planets' gas envelopes, we consider a scenario wherein they formed by core-nucleated accretion inside the snow-line and consist of an Earth-like composition rocky core (32\% by mass iron and 68\% by mass silicate) surrounded by an H/He envelope. For a given point in planet mass-radius-incident flux parameter space, planet interior structure models are used to constrain the distribution of each planet's mass between the H/He envelope and the heavy element interior. We model each planet's interior structure following an approach similar to \citet{Rogers2011}, but with updated opacities from \citet{Valencia2013}. An envelope metallicity of $30\times$ solar and a Bond albedo of 0.2 are assumed. We sample the posterior distribution on the envelope mass fraction by 1) drawing $10^5$ samples from the planet mass-radius-flux posterior distribution returned by the orbital, stellar, and TTV MCMC fits, and 2) computing the envelope mass fraction for each sample from the planet interior structure models. 

\subsection{PH3~b}

With the large relative uncertainty in its measured mass, the bulk compositions that are consistent with the measured properties of PH3~b encompass a wide range of possibilities, including rocky planet, water-planet, and H/He-enshrouded rocky core scenarios. PH3~b could be a rocky planet with its transit radius defined by a rocky surface; integrating the posterior probability distribution on PH3b's mass, radius, and incident flux, we find a 16\% posterior probability that PH~3b is more dense than a pure silicate sphere. Most of the posterior probability on PH3~b's properties, however, falls in a low-density regime in which the planet must have a volatile envelope comprised of some a combination of astrophysical ices (dominated by water), hydrogen, and helium.  At the low-mass extreme, there is a 12\% posterior probability that PH3~b requires a hydrogen-dominated envelope and is less dense than a pure H$_2$O sphere, but scenarios where the planet consists of a mixture of rock and water are viable at intermediate masses (and notably at masses near the best fit). In the scenario where PH3~b has a rocky Earth-like composition heavy element interior surrounded by H/He, the H/He mass fraction of PH3~b is determined to be $0.24^{+0.28}_{-0.22}\%$ (see Figure~\ref{fig:bc}) or $\Delta R_{\rm{env}}/R_{\rm{p}} = 22\pm14\%$ ($\Delta R_{\rm{env}} = 0.47^{+0.32}_{-0.30}~R_{\oplus}$).

\subsection{PH3~c}

\begin{figure*}[tbp]
	\centering
		\includegraphics[width=1.00\textwidth]{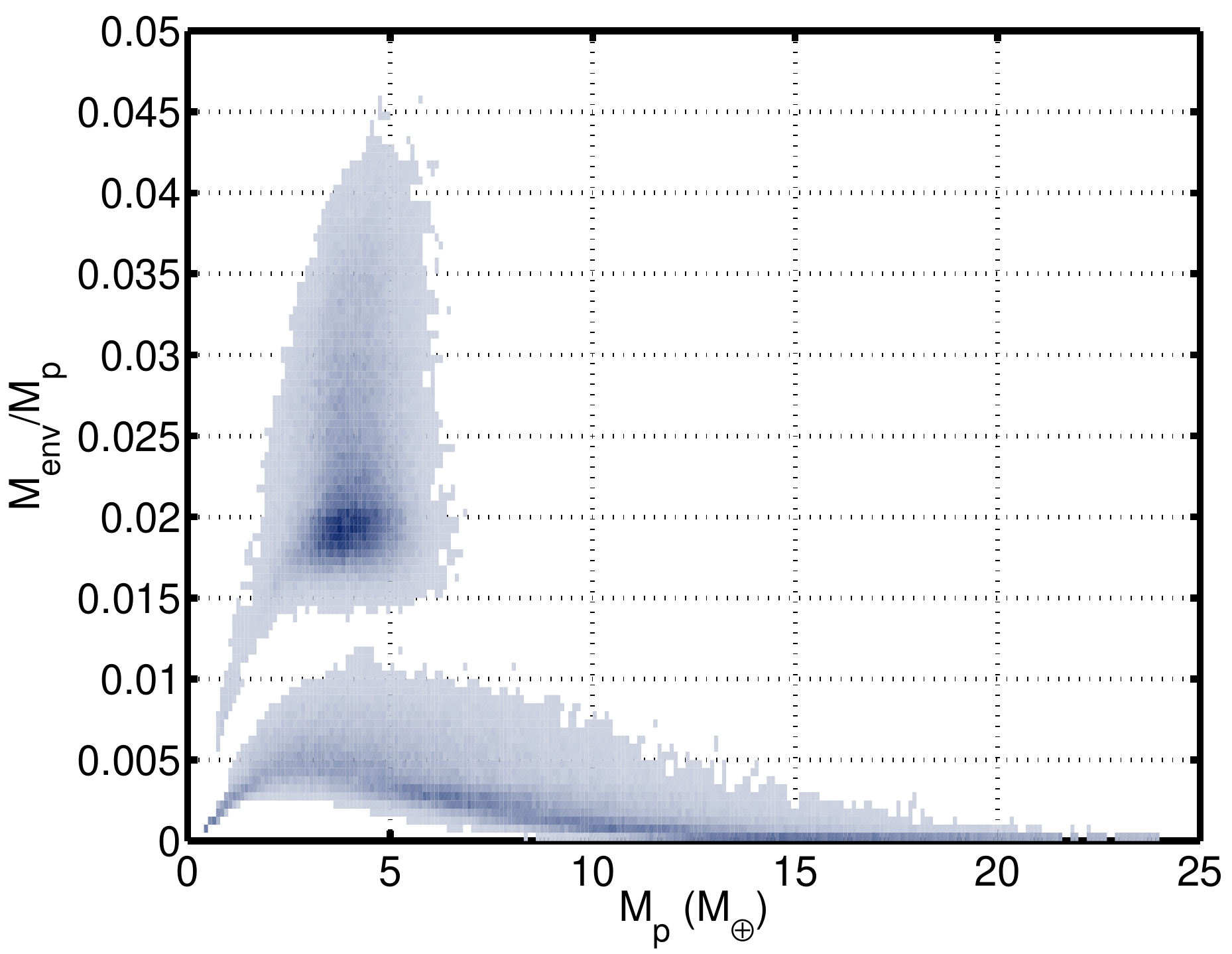}
	\caption{The posterior probability density distribution for PH3~b (top) and c (bottom), with darker blue representing a relatively higher probability, as a function of planet mass, $M_{\rm{P}}$ and the ratio of envelope mass to planetary mass, $M_{\rm{env}}/M_{\rm{P}}$. For both planets, we assume the cores are composed of heavy elements with no ices. PH3~c requires a significant envelope contribution to the total mass, while PH3~c's mass uncertainty leads to a broad range of possible compositions.}
	\label{fig:bc}
\end{figure*}

We find that PH3~c is $2.1^{+0.8}_{-0.3}\%$ by mass H/He (see Figure~\ref{fig:bc}), assuming the planet has a rocky core with no ices.  This corresponds to a radial extent of the H/He envelope of $\Delta R_{\rm{env}}/R_{\rm{p}} = 49^{+5}_{-4}\%$, or $\Delta R_{\rm{env}} = 1.31^{+0.32}_{-0.15}~R_{\oplus}$.  PH3~c must have a hydrogen-dominated envelope of light gases; there is less than $\sim~0.001\%$ posterior probability that PH3~c is equal density to or more dense than a pure H$_2$O body. 

\citet{Lopez2013} also explore the mass-radius relationship as a function of incident stellar flux, age, and composition.  They find that, at fixed age and flux, the radius acts as a proxy for the composition of Neptune-like planets, such as PH3~c ($R=2.68\pm0.17R_\oplus$).  PH3~c conveniently falls almost directly on top of one of their provided grid points, (mass, incident flux, age) $=$ ($3.6M_\oplus$, $10S_\oplus$, 1 Gyr), whereas our derived values are ($4.0\pm0.9M_\oplus$, $10.7\pm1.8S_\oplus$, $0.65\pm0.44$ Gyr (model fit) or $1.0\pm0.3$ Gyr (gyrochronology)).  \citet{Lopez2013} calculates the radius for two metallicities:  solar and $50\times$ solar.  PH3 has a metallicity of $\rm{[Fe/H]}=0.05\pm0.04$, although the atmospheres of Neptunes may be significantly enhanced in metal \citep{Fortney2013,Morley2013}.  A quadratic interpolation of their grid at ($3.6M_\oplus$, $10S_\oplus$, 1 Gyr) finds that the H/He mass fraction is 3.2\% for solar metallicity and 2.5\% for the enhanced metallicity model, which is consistent with the models we have computed.  

The results that PH3~c has a couple percent by mass H/He (and that PH3~b has no more than a few tenths of a percent by mass H/He) are robust. However, errors in the assumed planet energy budget, heavy element interior composition, envelope metallicity, and albedo could lead to small systematic shifts in the quantitative values of the H/He mass fractions quoted.

\subsection{PH3~d}

\begin{figure*}[tbp]
	\centering
		\includegraphics[width=1.00\textwidth]{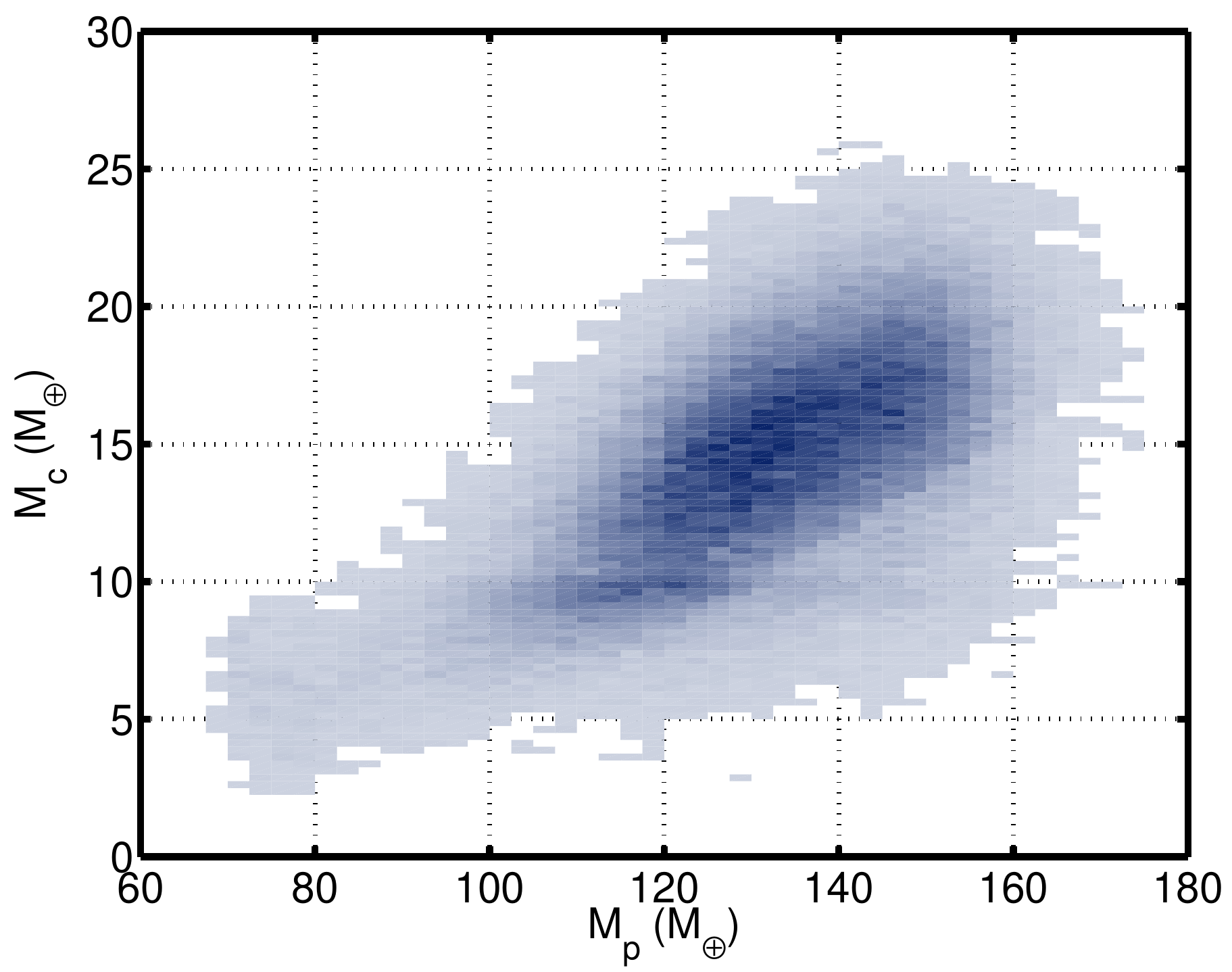}
	\caption{The posterior probability density distribution of PH3~d's core mass as a function of its total mass.}
	\label{fig:d}
\end{figure*}

PH3~d lies in the Jovian planet regime; its composition is dominated by hydrogen and helium. Interpolating among model grids from 1~Gyr-old Jovian planets from \citet{Fortney2007}, we compute the planet core mass for $10^5$ draws from the mass-radius-flux posterior distribution of PH3~d. In this way, we estimate a heavy-element core mass of $14\pm4~M_{\oplus}$ (see Figure~\ref{fig:d}), corresponding to an envelope mass fraction of $89^{+0.03}_{-0.02}\%$. We note that \citet{Fortney2007} consider fully differentiated planet with cores that are 50\% ice and 50\% rock by mass. If the heavy elements are distributed through the planet envelope or are made up of a different mixture of ice and rock, the inferred planet core mass will be affected.

\section{Discussion}

PH3~c avoided proper detection by other transit search routines, very likely due to the bias against detecting planets with large amplitude TTVs \citep{Garcia-Melendo2011}.  This may be the reason why the \textit{Kepler} pipeline misidentified this planet as a TCE with a period five times too large.  The Quasiperiodic Automated Search (QATS) algorithm \citep{Carter2013}, designed for detecting planets with TTVs, originally missed PH3~c as well due to PH3's strong stellar variability.  However, an upgraded version of QATS can now successfully identify this planet with improved detrending (E. Kruse 2014, private communication).  

It is curious that both planet pairs (outer/middle and middle/inner) have very near the same period ratio ($\sim1.91$), with $|1-(P_{\rm{out}}/P_{\rm{mid}})/(P_{\rm{mid}}/P_{\rm{in}})| = 0.385\%$.  There are 127 stars with $3+$ confirmed planets, in which there are 186 unique sets of three consecutive planets\footnote{http://exoplanetarchive.ipac.caltech.edu, last accessed June 19, 2014} \citep{Wright2011}.  Only four have the same ratio of period ratios to within 0.385\%.  The Laplace resonant GJ 876 is one of these, as is the Laplace resonant candidate system HR 8799, although it is a directly imaged planetary system with large error bars in period.  The other two are Kepler-207 and Kepler-229.  In the 174 KOIs with $3+$ confirmed candidates, there are 242 unique sets of three consecutive planets. Six of these have the same period ratio to within 0.385\%:  KOIs 665, 757, 869, 1151, 1358, and 2693.  

\begin{figure*}[tbp]
	\centering
		\includegraphics[bb=160 0 410 700,scale=0.9]{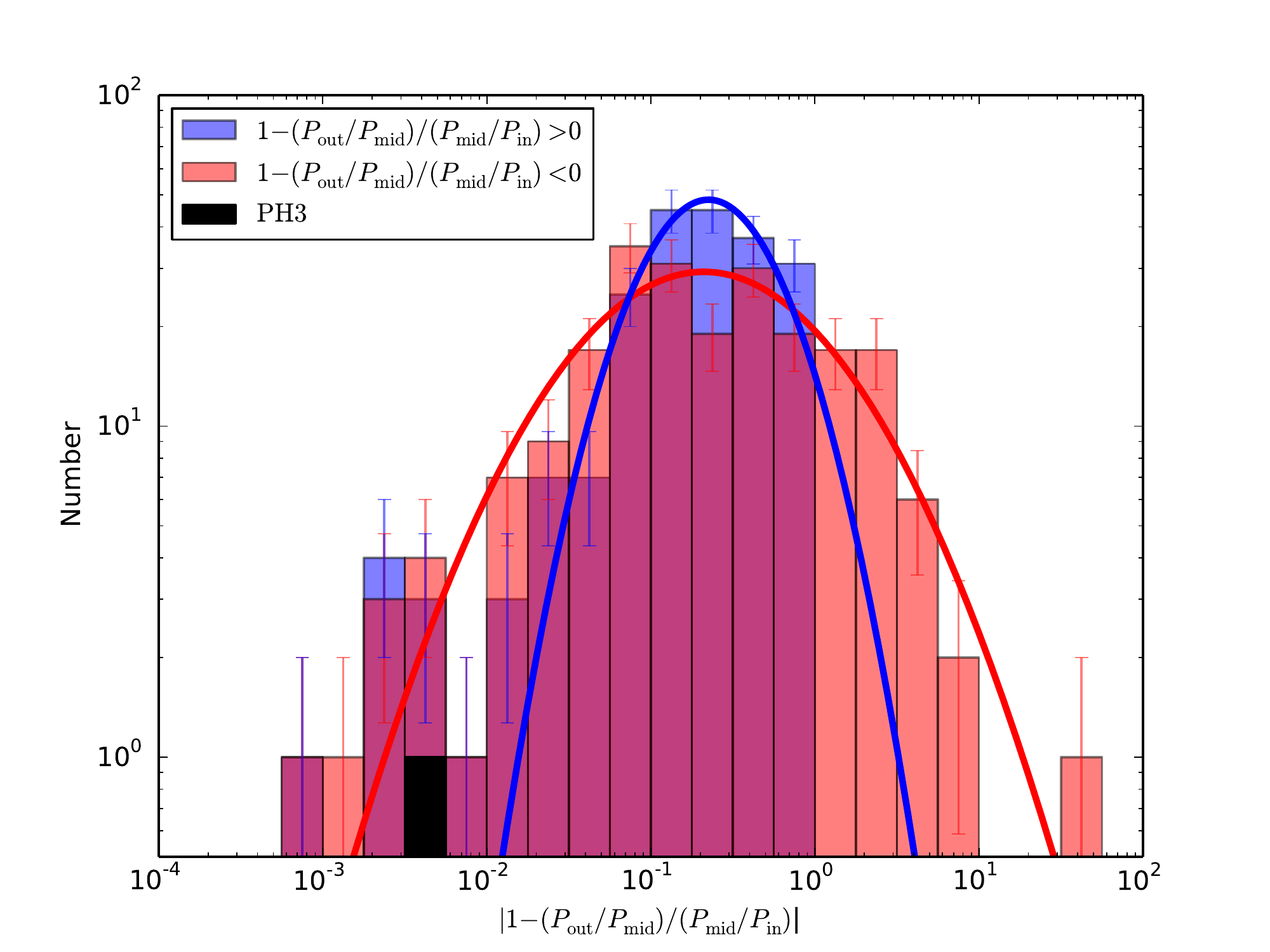}
	\caption{The difference of one minus the ratio of period ratios for consecutive sets of three planets for the combined set of both \textit{Kepler} candidates and confirmed planets.  The blue histogram represents the three-planet systems where $P_{\rm{out}}/P_{\rm{mid}} < P_{\rm{mid}}/P_{\rm{in}}$, and the red histogram represents the planets where $P_{\rm{out}}/P_{\rm{mid}} > P_{\rm{mid}}/P_{\rm{in}}$, where the value of the red population is multiplied by $-1$ in order to compare to the blue population.  The purple represents overlapping red and blue populations, while the black histogram represents PH3, which is also included in the blue population.  The blue population drops off at high values of the abscissa due to the mathematical truncation at 1.0 as $(P_{\rm{out}}/P_{\rm{mid}})/(P_{\rm{mid}}/P_{\rm{in}})$ approaches 0. The two lines represent best fits for a log-normal distribution. The log-normal fit fails for the blue population as there exists an excess of planet triplets near zero.}
	\label{fig:perrat}
\end{figure*}

For the half of the distribution of the relative difference in the pair-wise period ratios that is not mathematically truncated at 1.0 (see the red histogram Figure~\ref{fig:perrat}), a log-normal fit can be reasonably applied to the distribution.  A variable with a log-normal distribution indicates that the variable is the product of many independent, random variables.  However, in the blue population, which includes PH3, a log-normal does not fit the allowed region of parameter space ($<1$).  There is an excess of planet triplets near zero.  This implies that the two populations are affected by different processes.  However, any further analysis of this is beyond the scope of this paper and is left for future studies.

As of now, there is only one confirmed exoplanet system in a Laplace resonance ($\sim$ 4:2:1 MMR):  GJ 876 \citep{Rivera2010,Marti2013}.  GJ 876 b has a period of 61.12 days \citep{Marcy1998,Delfosse1998}, GJ 876 c has a period of 30.09 days \citep{Marcy2001}, and GJ 876 e orbits every 124.26 days \citep{Rivera2010}, while the period ratios of outer/middle and middle/inner of these three planets are the same to three digits, 2.03.  The Laplace resonance of GJ 876 was further explored by \citet{Marti2013}, who concluded that GJ 876's Laplace resonant solutions are stable, but surrounded by extremely unstable regions.  HD 8799 has three confirmed planets \citep{Marois2008} and also has regions of parameter space that are stable to the 4:2:1 MMR \citep{Reidemeister2009}.  

\begin{figure*}[tbp]
	\centering
		\includegraphics[bb=160 00 410 700,scale=0.9]{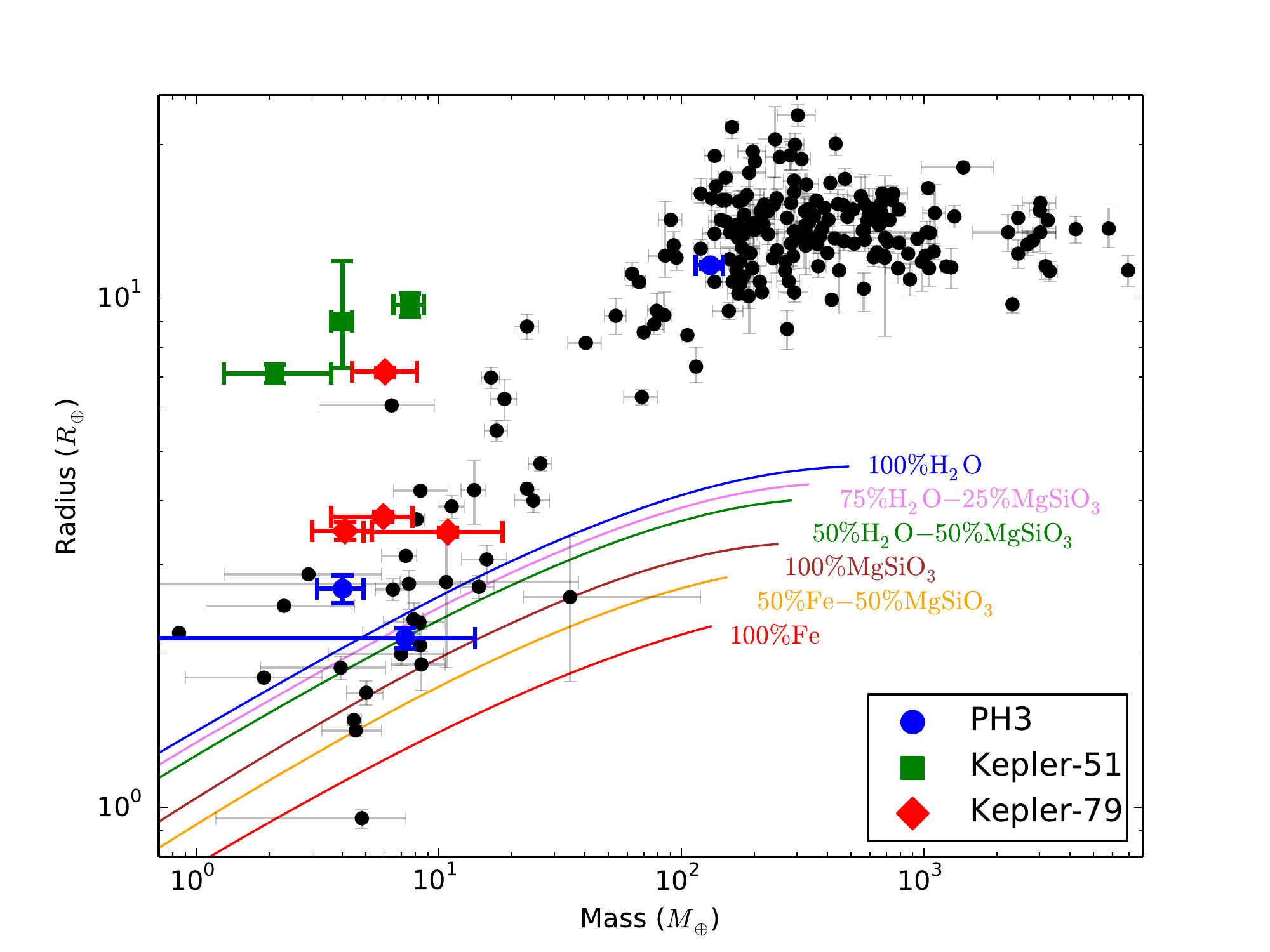}
	\caption{The mass-radius diagram for all planets with measured masses, radii, and error bars from \url{http://exoplanets.org/} \citep{Wright2011} in black circles.  Blue circles represent PH3 candidates, green squares Kepler-51 \citep{Masuda2014}, and red diamonds Kepler-79 \citep{Jontof-Hutter2014}.  The colored lines represent the masses and radii of planets in different two-component models of planetary compositions from \citet{Zeng2013} as indicated on the right.  The 100\% water-ice model does not include any steam atmosphere that would occur for warm/hot water-ice planets.}
	\label{fig:mr}
\end{figure*}

In Figure~\ref{fig:mr}, we show the mass-radius diagram of PH3's planets compared to other confirmed planets (those with masses, radii, and errors from \url{http://exoplanets.org/} \citep{Wright2011}), two recent notable results discussed below, and mass-radius-composition models provided by \citet{Zeng2013}.  

With a density of 1.2 g/cm$^3$, PH3~c joins a growing population of low-mass ($M<10M_\oplus$), low-density planets that require significant H/He gaseous atmospheres, e.g., Kepler-11 \citep[$\rho=1.7$, 1.28, 0.66, 0.58, and 0.69 g/cm$^3$ for Kepler-11~b, c, d, e, and f, respectively,][]{Lissauer2013}, Kepler-87c \citep[$\rho=0.152$ g/cm$^3$,][]{Ofir2014}, GJ 1214~b \citep[$\rho=1.87$ g/cm$^3$,][]{Charbonneau2009}, and Kepler-36c \citep[$\rho=0.89$ g/cm$^3$,][]{Carter2012}.  In the Kepler-79 system \citep{Xie2014}, \citet{Jontof-Hutter2014} calculate more precise masses of its four planets via TTVs and finds all four with low densities.  The least dense, Kepler-79d, has $\rho=0.09$ g/cm$^3$ with a mass of just $6.0M_\oplus$. Kepler-51 is even more extreme with two confirmed planets \citep{Steffen2012} both with densities of 0.03 g/cm$^3$ and masses of $2.1M_\oplus$ and $4.0M_\oplus$ \citep{Masuda2014}.  The third planet in the system confirmed by \citet{Masuda2014} has $\rho=0.05$ g/cm$^3$ and $M=7.6M_\oplus$. 

The dominant method for discovering these low-mass, low-density planets is via TTV analysis.  However, TTV discoveries may be biased to low density results since, at constant mass, planets with larger radius and thus lower density will be detected more readily and with more precise midtransit times \citep{Jontof-Hutter2014}. The likelihood of transit detection approximately increases with $R^2_{\rm{P}}$.  Therefore, although more massive planets would create larger, more detectable TTV signals, the balance is likely in favor of a bias towards discovering low density planets.

As noted in \citet{Masuda2014}, a number of low-density planets are found near MMR.  However, TTV detections are best-suited for planets near MMR as they provide the clearest and largest amplitude signals, so this feature may be a selection effect.  More discoveries of low-mass, low-density planets and future theoretical work examining this new population of low-mass, low density planets will be needed to address whether this is a truly a selection effect or a signature of planet formation or migration.

\noindent\textit{Acknowledgements} 

The data presented in this paper are the result of the efforts of the PH volunteers, without whom this work would not have been possible. Their contributions are individually acknowledged at \url{http://www.planethunters.org/authors}. The authors thank the  PH volunteers who participated in identifying and analyzing the candidates presented in this paper.

E.A. acknowledges funding by NSF Career grant AST 0645416; NASA Astrobiology Institutes Virtual Planetary Laboratory, supported by NASA under cooperative agreement NNH05ZDA001C; and NASA Origins of Solar Systems grant 12-OSS12-0011. K.M.D. acknowledges the support of an NSF Graduate Research Fellowship.  DF acknowledges funding support for PlanetHunters.org from Yale University and support from the NASA Supplemental Outreach Award, 10-OUTRCH.210-0001 and the NASA ADAP12-0172. LAR gratefully acknowledges support provided by NASA through Hubble Fellowship grant \#HF-51313 awarded by the Space Telescope Science Institute, which is operated by the Association of Universities for Research in Astronomy, Inc., for NASA, under contract NAS 5-26555.  KS gratefully acknowledges support from Swiss National Science Foundation Grant PP00P2\_138979/1. MES is supported in part by an Academia Sinica postdoctoral fellowship. 

This research has made use of the Exoplanet Orbit Database and the Exoplanet Data Explorer at exoplanets.org. The Zooniverse is supported by The Leverhulme Trust and by the Alfred P. Sloan foundation. PH is supported in part by NASA JPL's PlanetQuest program. The Talk system used by PH was built during work supported by the NSF under Grant No. DRL-0941610. We gratefully acknowledge the dedication and achievements of \textit{Kepler} science team and all those who contributed to the success of the mission. We acknowledge use of public release data served by the NASA/IPAC/NExScI Star and Exoplanet Database, which is operated by the Jet Propulsion Laboratory, California Institute of Technology, under contract with the National Aeronautics and Space Administration. This research has made use of NASA's Astrophysics Data System Bibliographic Services. This paper includes data collected by the \textit{Kepler} spacecraft, and we gratefully acknowledge the entire \textit{Kepler} mission team's efforts in obtaining and providing the light curves used in this analysis. Funding for the \textit{Kepler} mission is provided by the NASA Science Mission directorate. Support for MAST for non-HST data is provided by the NASA Office of Space Science via grant NNX13AC07G and by other grants and contracts.  Some of the data presented in this paper were obtained from the Mikulski Archive for Space Telescopes (MAST). STScI is operated by the Association of Universities for Research in Astronomy, Inc., under NASA contract NAS5-26555. Support for MAST for non-HST data is provided by the NASA Office of Space Science via grant NNX13AC07G and by other grants and contracts.  The data presented herein were partly obtained at the W. M. Keck Observatory, which is op-erated as a scientific partnership among the California Institute of Technology, the University of California, and the National Aeronautics and Space Administration. The Observatory was made possible by the generous financial support of the W. M. Keck Foundation.


\begin{thebibliography}{90}
\expandafter\ifx\csname natexlab\endcsname\relax\def\natexlab#1{#1}\fi

\bibitem[{{Abazajian} {et~al.}(2009){Abazajian}, {Adelman-McCarthy},
  {Ag{\"u}eros}, {Allam}, {Allende Prieto}, {An}, {Anderson}, {Anderson},
  {Annis}, {Bahcall}, \& et~al.}]{Abazajian2009}
{Abazajian}, K.~N., {et~al.} 2009, \apjs, 182, 543

\bibitem[{{Agol} {et~al.}(2005){Agol}, {Steffen}, {Sari}, \&
  {Clarkson}}]{Agol2005}
{Agol}, E., {Steffen}, J., {Sari}, R., \& {Clarkson}, W. 2005, \mnras, 359, 567

\bibitem[{{Aksnes}(1988)}]{Aksnes1988}
{Aksnes}, K. 1988, in Long-term Dynamical Behaviour of Natural and Artificial
  N-body Systems, ed. A.~D. {Roy}, 125--139

\bibitem[{{Ballard} {et~al.}(2011){Ballard}, {Fabrycky}, {Fressin},
  {Charbonneau}, {Desert}, {Torres}, {Marcy}, {Burke}, {Isaacson}, {Henze},
  {Steffen}, {Ciardi}, {Howell}, {Cochran}, {Endl}, {Bryson}, {Rowe}, {Holman},
  {Lissauer}, {Jenkins}, {Still}, {Ford}, {Christiansen}, {Middour}, {Haas},
  {Li}, {Hall}, {McCauliff}, {Batalha}, {Koch}, \& {Borucki}}]{Ballard2011}
{Ballard}, S., {et~al.} 2011, \apj, 743, 200

\bibitem[{{Barnes}(2007)}]{Barnes2007}
{Barnes}, S.~A. 2007, \apj, 669, 1167

\bibitem[{{Batalha} {et~al.}(2013){Batalha}, {Rowe}, {Bryson}, {Barclay},
  {Burke}, {Caldwell}, {Christiansen}, {Mullally}, {Thompson}, {Brown},
  {Dupree}, {Fabrycky}, {Ford}, {Fortney}, {Gilliland}, {Isaacson}, {Latham},
  {Marcy}, {Quinn}, {Ragozzine}, {Shporer}, {Borucki}, {Ciardi}, {Gautier},
  {Haas}, {Jenkins}, {Koch}, {Lissauer}, {Rapin}, {Basri}, {Boss}, {Buchhave},
  {Carter}, {Charbonneau}, {Christensen-Dalsgaard}, {Clarke}, {Cochran},
  {Demory}, {Desert}, {Devore}, {Doyle}, {Esquerdo}, {Everett}, {Fressin},
  {Geary}, {Girouard}, {Gould}, {Hall}, {Holman}, {Howard}, {Howell},
  {Ibrahim}, {Kinemuchi}, {Kjeldsen}, {Klaus}, {Li}, {Lucas}, {Meibom},
  {Morris}, {Pr{\v s}a}, {Quintana}, {Sanderfer}, {Sasselov}, {Seader},
  {Smith}, {Steffen}, {Still}, {Stumpe}, {Tarter}, {Tenenbaum}, {Torres},
  {Twicken}, {Uddin}, {Van Cleve}, {Walkowicz}, \& {Welsh}}]{Batalha2013}
{Batalha}, N.~M., {et~al.} 2013, \apjs, 204, 24

\bibitem[{{Borucki} {et~al.}(2011){Borucki}, {Koch}, {Basri}, {Batalha},
  {Brown}, {Bryson}, {Caldwell}, {Christensen-Dalsgaard}, {Cochran}, {DeVore},
  {Dunham}, {Gautier}, {Geary}, {Gilliland}, {Gould}, {Howell}, {Jenkins},
  {Latham}, {Lissauer}, {Marcy}, {Rowe}, {Sasselov}, {Boss}, {Charbonneau},
  {Ciardi}, {Doyle}, {Dupree}, {Ford}, {Fortney}, {Holman}, {Seager},
  {Steffen}, {Tarter}, {Welsh}, {Allen}, {Buchhave}, {Christiansen}, {Clarke},
  {Das}, {D{\'e}sert}, {Endl}, {Fabrycky}, {Fressin}, {Haas}, {Horch},
  {Howard}, {Isaacson}, {Kjeldsen}, {Kolodziejczak}, {Kulesa}, {Li}, {Lucas},
  {Machalek}, {McCarthy}, {MacQueen}, {Meibom}, {Miquel}, {Prsa}, {Quinn},
  {Quintana}, {Ragozzine}, {Sherry}, {Shporer}, {Tenenbaum}, {Torres},
  {Twicken}, {Van Cleve}, {Walkowicz}, {Witteborn}, \& {Still}}]{Borucki2011}
{Borucki}, W.~J., {et~al.} 2011, \apj, 736, 19

\bibitem[{{Bressan} {et~al.}(2012){Bressan}, {Marigo}, {Girardi}, {Salasnich},
  {Dal Cero}, {Rubele}, \& {Nanni}}]{Bressan2012}
{Bressan}, A., {Marigo}, P., {Girardi}, L., {Salasnich}, B., {Dal Cero}, C.,
  {Rubele}, S., \& {Nanni}, A. 2012, \mnras, 427, 127

\bibitem[{{Burke} {et~al.}(2014){Burke}, {Bryson}, {Mullally}, {Rowe},
  {Christiansen}, {Thompson}, {Coughlin}, {Haas}, {Batalha}, {Caldwell},
  {Jenkins}, {Still}, {Barclay}, {Borucki}, {Chaplin}, {Ciardi}, {Clarke},
  {Cochran}, {Demory}, {Esquerdo}, {Gautier}, {Gilliland}, {Girouard}, {Havel},
  {Henze}, {Howell}, {Huber}, {Latham}, {Li}, {Morehead}, {Morton}, {Pepper},
  {Quintana}, {Ragozzine}, {Seader}, {Shah}, {Shporer}, {Tenenbaum}, {Twicken},
  \& {Wolfgang}}]{Burke2014}
{Burke}, C.~J., {et~al.} 2014, \apjs, 210, 19

\bibitem[{{Cabrera} {et~al.}(2014){Cabrera}, {Csizmadia}, {Lehmann}, {Dvorak},
  {Gandolfi}, {Rauer}, {Erikson}, {Dreyer}, {Eigm{\"u}ller}, \&
  {Hatzes}}]{Cabrera2014}
{Cabrera}, J., {et~al.} 2014, \apj, 781, 18

\bibitem[{{Carter} \& {Agol}(2013)}]{Carter2013}
{Carter}, J.~A., \& {Agol}, E. 2013, \apj, 765, 132

\bibitem[{{Carter} \& {Winn}(2009)}]{Carter2009}
{Carter}, J.~A., \& {Winn}, J.~N. 2009, \apj, 704, 51

\bibitem[{{Carter} {et~al.}(2012){Carter}, {Agol}, {Chaplin}, {Basu},
  {Bedding}, {Buchhave}, {Christensen-Dalsgaard}, {Deck}, {Elsworth},
  {Fabrycky}, {Ford}, {Fortney}, {Hale}, {Handberg}, {Hekker}, {Holman},
  {Huber}, {Karoff}, {Kawaler}, {Kjeldsen}, {Lissauer}, {Lopez}, {Lund},
  {Lundkvist}, {Metcalfe}, {Miglio}, {Rogers}, {Stello}, {Borucki}, {Bryson},
  {Christiansen}, {Cochran}, {Geary}, {Gilliland}, {Haas}, {Hall}, {Howard},
  {Jenkins}, {Klaus}, {Koch}, {Latham}, {MacQueen}, {Sasselov}, {Steffen},
  {Twicken}, \& {Winn}}]{Carter2012}
{Carter}, J.~A., {et~al.} 2012, Science, 337, 556

\bibitem[{{Charbonneau} {et~al.}(2009){Charbonneau}, {Berta}, {Irwin}, {Burke},
  {Nutzman}, {Buchhave}, {Lovis}, {Bonfils}, {Latham}, {Udry}, {Murray-Clay},
  {Holman}, {Falco}, {Winn}, {Queloz}, {Pepe}, {Mayor}, {Delfosse}, \&
  {Forveille}}]{Charbonneau2009}
{Charbonneau}, D., {et~al.} 2009, \nat, 462, 891

\bibitem[{{Cutri} \& {et al.}(2012)}]{Cutri2012}
{Cutri}, R.~M., \& {et al.} 2012, VizieR Online Data Catalog, 2311, 0

\bibitem[{{Cutri} {et~al.}(2003){Cutri}, {Skrutskie}, {van Dyk}, {Beichman},
  {Carpenter}, {Chester}, {Cambresy}, {Evans}, {Fowler}, {Gizis}, {Howard},
  {Huchra}, {Jarrett}, {Kopan}, {Kirkpatrick}, {Light}, {Marsh}, {McCallon},
  {Schneider}, {Stiening}, {Sykes}, {Weinberg}, {Wheaton}, {Wheelock}, \&
  {Zacarias}}]{Cutri2003}
{Cutri}, R.~M., {et~al.} 2003, VizieR Online Data Catalog, 2246, 0

\bibitem[{{Deck} {et~al.}(2014){Deck}, {Agol}, {Holman}, \&
  {Nesvorn{\'y}}}]{Deck2014}
{Deck}, K.~M., {Agol}, E., {Holman}, M.~J., \& {Nesvorn{\'y}}, D. 2014, \apj,
  787, 132

\bibitem[{{Delfosse} {et~al.}(1998){Delfosse}, {Forveille}, {Mayor}, {Perrier},
  {Naef}, \& {Queloz}}]{Delfosse1998}
{Delfosse}, X., {Forveille}, T., {Mayor}, M., {Perrier}, C., {Naef}, D., \&
  {Queloz}, D. 1998, \aap, 338, L67

\bibitem[{{Dotter} {et~al.}(2008){Dotter}, {Chaboyer}, {Jevremovi{\'c}},
  {Kostov}, {Baron}, \& {Ferguson}}]{Dotter2008}
{Dotter}, A., {Chaboyer}, B., {Jevremovi{\'c}}, D., {Kostov}, V., {Baron}, E.,
  \& {Ferguson}, J.~W. 2008, \apjs, 178, 89

\bibitem[{{Eastman} {et~al.}(2013){Eastman}, {Gaudi}, \& {Agol}}]{Eastman2013}
{Eastman}, J., {Gaudi}, B.~S., \& {Agol}, E. 2013, \pasp, 125, 83

\bibitem[{{Fabrycky} {et~al.}(2012){Fabrycky}, {Ford}, {Steffen}, {Rowe},
  {Carter}, {Moorhead}, {Batalha}, {Borucki}, {Bryson}, {Buchhave},
  {Christiansen}, {Ciardi}, {Cochran}, {Endl}, {Fanelli}, {Fischer}, {Fressin},
  {Geary}, {Haas}, {Hall}, {Holman}, {Jenkins}, {Koch}, {Latham}, {Li},
  {Lissauer}, {Lucas}, {Marcy}, {Mazeh}, {McCauliff}, {Quinn}, {Ragozzine},
  {Sasselov}, \& {Shporer}}]{Fabrycky2012b}
{Fabrycky}, D.~C., {et~al.} 2012, \apj, 750, 114

\bibitem[{{Fabrycky} {et~al.}(2014){Fabrycky}, {Lissauer}, {Ragozzine}, {Rowe},
  {Steffen}, {Agol}, {Barclay}, {Batalha}, {Borucki}, {Ciardi}, {Ford},
  {Gautier}, {Geary}, {Holman}, {Jenkins}, {Li}, {Morehead}, {Morris},
  {Shporer}, {Smith}, {Still}, \& {Van Cleve}}]{Fabrycky2012a}
---. 2014, \apj, 790, 146

\bibitem[{{Fischer} {et~al.}(2012){Fischer}, {Schwamb}, {Schawinski},
  {Lintott}, {Brewer}, {Giguere}, {Lynn}, {Parrish}, {Sartori}, {Simpson},
  {Smith}, {Spronck}, {Batalha}, {Rowe}, {Jenkins}, {Bryson}, {Prsa},
  {Tenenbaum}, {Crepp}, {Morton}, {Howard}, {Beleu}, {Kaplan}, {Vannispen},
  {Sharzer}, {Defouw}, {Hajduk}, {Neal}, {Nemec}, {Schuepbach}, \&
  {Zimmermann}}]{Fischer2012}
{Fischer}, D.~A., {et~al.} 2012, \mnras, 419, 2900

\bibitem[{{Ford}(2006)}]{Ford2006}
{Ford}, E.~B. 2006, \apj, 642, 505

\bibitem[{{Ford} \& {Holman}(2007)}]{Ford2007}
{Ford}, E.~B., \& {Holman}, M.~J. 2007, \apjl, 664, L51

\bibitem[{{Foreman-Mackey} {et~al.}(2013){Foreman-Mackey}, {Hogg}, {Lang}, \&
  {Goodman}}]{Foreman-Mackey2013}
{Foreman-Mackey}, D., {Hogg}, D.~W., {Lang}, D., \& {Goodman}, J. 2013, \pasp,
  125, 306

\bibitem[{{Fortney} {et~al.}(2007){Fortney}, {Marley}, \&
  {Barnes}}]{Fortney2007}
{Fortney}, J.~J., {Marley}, M.~S., \& {Barnes}, J.~W. 2007, \apj, 659, 1661

\bibitem[{{Fortney} {et~al.}(2013){Fortney}, {Mordasini}, {Nettelmann},
  {Kempton}, {Greene}, \& {Zahnle}}]{Fortney2013}
{Fortney}, J.~J., {Mordasini}, C., {Nettelmann}, N., {Kempton}, E.~M.-R.,
  {Greene}, T.~P., \& {Zahnle}, K. 2013, \apj, 775, 80

\bibitem[{{Fortson} {et~al.}(2012){Fortson}, {Masters}, {Nichol}, {Borne},
  {Edmondson}, {Lintott}, {Raddick}, {Schawinski}, \& {Wallin}}]{Fortson2012}
{Fortson}, L., {et~al.} 2012, in Advances in Machine Learning and Data Mining
  for Astronomy, CRC Press, Taylor {\&} Francis Group, Eds.: Michael J.~Way,
  Jeffrey D.~Scargle, Kamal M.~Ali, Ashok N.~Srivastava, p.~213-236, ed. M.~J.
  {Way}, J.~D. {Scargle}, K.~M. {Ali}, \& A.~N. {Srivastava}, 213--236

\bibitem[{{Fressin} {et~al.}(2012){Fressin}, {Torres}, {Rowe}, {Charbonneau},
  {Rogers}, {Ballard}, {Batalha}, {Borucki}, {Bryson}, {Buchhave}, {Ciardi},
  {D{\'e}sert}, {Dressing}, {Fabrycky}, {Ford}, {Gautier}, {Henze}, {Holman},
  {Howard}, {Howell}, {Jenkins}, {Koch}, {Latham}, {Lissauer}, {Marcy},
  {Quinn}, {Ragozzine}, {Sasselov}, {Seager}, {Barclay}, {Mullally}, {Seader},
  {Still}, {Twicken}, {Thompson}, \& {Uddin}}]{Fressin2012}
{Fressin}, F., {et~al.} 2012, \nat, 482, 195

\bibitem[{{Garc{\'{\i}}a-Melendo} \&
  {L{\'o}pez-Morales}(2011)}]{Garcia-Melendo2011}
{Garc{\'{\i}}a-Melendo}, E., \& {L{\'o}pez-Morales}, M. 2011, \mnras, 417, L16

\bibitem[{{Gazak} {et~al.}(2012){Gazak}, {Johnson}, {Tonry}, {Dragomir},
  {Eastman}, {Mann}, \& {Agol}}]{Gazak2012}
{Gazak}, J.~Z., {Johnson}, J.~A., {Tonry}, J., {Dragomir}, D., {Eastman}, J.,
  {Mann}, A.~W., \& {Agol}, E. 2012, Advances in Astronomy, 2012

\bibitem[{{Gladman}(1993)}]{Gladman1993}
{Gladman}, B. 1993, Icarus, 106, 247

\bibitem[{{Goodman} \& {Weare}(2010)}]{Goodman2010}
{Goodman}, J., \& {Weare}, J. 2010, Comm. App. Math. Comp. Sci., 5, 65

\bibitem[{{Holman} \& {Murray}(2005)}]{Holman2005}
{Holman}, M.~J., \& {Murray}, N.~W. 2005, Science, 307, 1288

\bibitem[{{Holman} {et~al.}(2010){Holman}, {Fabrycky}, {Ragozzine}, {Ford},
  {Steffen}, {Welsh}, {Lissauer}, {Latham}, {Marcy}, {Walkowicz}, {Batalha},
  {Jenkins}, {Rowe}, {Cochran}, {Fressin}, {Torres}, {Buchhave}, {Sasselov},
  {Borucki}, {Koch}, {Basri}, {Brown}, {Caldwell}, {Charbonneau}, {Dunham},
  {Gautier}, {Geary}, {Gilliland}, {Haas}, {Howell}, {Ciardi}, {Endl},
  {Fischer}, {F{\"u}r{\'e}sz}, {Hartman}, {Isaacson}, {Johnson}, {MacQueen},
  {Moorhead}, {Morehead}, \& {Orosz}}]{Holman2010}
{Holman}, M.~J., {et~al.} 2010, Science, 330, 51

\bibitem[{{Huber} {et~al.}(2014){Huber}, {Silva Aguirre}, {Matthews},
  {Pinsonneault}, {Gaidos}, {Garc{\'{\i}}a}, {Hekker}, {Mathur}, {Mosser},
  {Torres}, {Bastien}, {Basu}, {Bedding}, {Chaplin}, {Demory}, {Fleming},
  {Guo}, {Mann}, {Rowe}, {Serenelli}, {Smith}, \& {Stello}}]{Huber2014}
{Huber}, D., {et~al.} 2014, \apjs, 211, 2

\bibitem[{{Jenkins} {et~al.}(2002){Jenkins}, {Caldwell}, \&
  {Borucki}}]{Jenkins2002}
{Jenkins}, J.~M., {Caldwell}, D.~A., \& {Borucki}, W.~J. 2002, \apj, 564, 495

\bibitem[{{Jenkins} {et~al.}(2010){Jenkins}, {Chandrasekaran}, {McCauliff},
  {Caldwell}, {Tenenbaum}, {Li}, {Klaus}, {Cote}, \& {Middour}}]{Jenkins2010}
{Jenkins}, J.~M., {et~al.} 2010, in Society of Photo-Optical Instrumentation
  Engineers (SPIE) Conference Series, Vol. 7740, Society of Photo-Optical
  Instrumentation Engineers (SPIE) Conference Series

\bibitem[{{Jontof-Hutter} {et~al.}(2014){Jontof-Hutter}, {Lissauer}, {Rowe}, \&
  {Fabrycky}}]{Jontof-Hutter2014}
{Jontof-Hutter}, D., {Lissauer}, J.~J., {Rowe}, J.~F., \& {Fabrycky}, D.~C.
  2014, \apj, 785, 15

\bibitem[{{Kipping}(2009)}]{Kipping2009}
{Kipping}, D.~M. 2009, \mnras, 392, 181

\bibitem[{{Kipping}(2013)}]{Kipping2013}
---. 2013, \mnras, 435, 2152

\bibitem[{{Lintott} {et~al.}(2011){Lintott}, {Schawinski}, {Bamford}, {Slosar},
  {Land}, {Thomas}, {Edmondson}, {Masters}, {Nichol}, {Raddick}, {Szalay},
  {Andreescu}, {Murray}, \& {Vandenberg}}]{Lintott2011}
{Lintott}, C., {et~al.} 2011, \mnras, 410, 166

\bibitem[{{Lintott} {et~al.}(2008){Lintott}, {Schawinski}, {Slosar}, {Land},
  {Bamford}, {Thomas}, {Raddick}, {Nichol}, {Szalay}, {Andreescu}, {Murray}, \&
  {Vandenberg}}]{Lintott2008}
{Lintott}, C.~J., {et~al.} 2008, \mnras, 389, 1179

\bibitem[{{Lintott} {et~al.}(2013){Lintott}, {Schwamb}, {Barclay}, {Sharzer},
  {Fischer}, {Brewer}, {Giguere}, {Lynn}, {Parrish}, {Batalha}, {Bryson},
  {Jenkins}, {Ragozzine}, {Rowe}, {Schwainski}, {Gagliano}, {Gilardi}, {Jek},
  {P{\"a}{\"a}kk{\"o}nen}, \& {Smits}}]{Lintott2013}
---. 2013, \aj, 145, 151

\bibitem[{{Lissauer} {et~al.}(2011){Lissauer}, {Fabrycky}, {Ford}, {Borucki},
  {Fressin}, {Marcy}, {Orosz}, {Rowe}, {Torres}, {Welsh}, {Batalha}, {Bryson},
  {Buchhave}, {Caldwell}, {Carter}, {Charbonneau}, {Christiansen}, {Cochran},
  {Desert}, {Dunham}, {Fanelli}, {Fortney}, {Gautier}, {Geary}, {Gilliland},
  {Haas}, {Hall}, {Holman}, {Koch}, {Latham}, {Lopez}, {McCauliff}, {Miller},
  {Morehead}, {Quintana}, {Ragozzine}, {Sasselov}, {Short}, \&
  {Steffen}}]{Lissauer2011}
{Lissauer}, J.~J., {et~al.} 2011, \nat, 470, 53

\bibitem[{{Lissauer} {et~al.}(2013){Lissauer}, {Jontof-Hutter}, {Rowe},
  {Fabrycky}, {Lopez}, {Agol}, {Marcy}, {Deck}, {Fischer}, {Fortney}, {Howell},
  {Isaacson}, {Jenkins}, {Kolbl}, {Sasselov}, {Short}, \&
  {Welsh}}]{Lissauer2013}
---. 2013, \apj, 770, 131

\bibitem[{{Lissauer} {et~al.}(2014){Lissauer}, {Marcy}, {Bryson}, {Rowe},
  {Jontof-Hutter}, {Agol}, {Borucki}, {Carter}, {Ford}, {Gilliland}, {Kolbl},
  {Star}, {Steffen}, \& {Torres}}]{Lissauer2014}
---. 2014, \apj, 784, 44

\bibitem[{{Lithwick} {et~al.}(2012){Lithwick}, {Xie}, \& {Wu}}]{Lithwick2012}
{Lithwick}, Y., {Xie}, J., \& {Wu}, Y. 2012, \apj, 761, 122

\bibitem[{{Lopez} \& {Fortney}(2014)}]{Lopez2013}
{Lopez}, E.~D., \& {Fortney}, J.~J. 2014, \apj, 792, 1

\bibitem[{{Marcy} {et~al.}(2001){Marcy}, {Butler}, {Fischer}, {Vogt},
  {Lissauer}, \& {Rivera}}]{Marcy2001}
{Marcy}, G.~W., {Butler}, R.~P., {Fischer}, D., {Vogt}, S.~S., {Lissauer},
  J.~J., \& {Rivera}, E.~J. 2001, \apj, 556, 296

\bibitem[{{Marcy} {et~al.}(1998){Marcy}, {Butler}, {Vogt}, {Fischer}, \&
  {Lissauer}}]{Marcy1998}
{Marcy}, G.~W., {Butler}, R.~P., {Vogt}, S.~S., {Fischer}, D., \& {Lissauer},
  J.~J. 1998, \apjl, 505, L147

\bibitem[{{Marois} {et~al.}(2008){Marois}, {Macintosh}, {Barman}, {Zuckerman},
  {Song}, {Patience}, {Lafreni{\`e}re}, \& {Doyon}}]{Marois2008}
{Marois}, C., {Macintosh}, B., {Barman}, T., {Zuckerman}, B., {Song}, I.,
  {Patience}, J., {Lafreni{\`e}re}, D., \& {Doyon}, R. 2008, Science, 322, 1348

\bibitem[{{Mart{\'{\i}}} {et~al.}(2013){Mart{\'{\i}}}, {Giuppone}, \&
  {Beaug{\'e}}}]{Marti2013}
{Mart{\'{\i}}}, J.~G., {Giuppone}, C.~A., \& {Beaug{\'e}}, C. 2013, \mnras,
  433, 928

\bibitem[{{Masuda}(2014)}]{Masuda2014}
{Masuda}, K. 2014, \apj, 783, 53

\bibitem[{{Mayor} \& {Queloz}(1995)}]{Mayor1995}
{Mayor}, M., \& {Queloz}, D. 1995, \nat, 378, 355

\bibitem[{{Mazeh} {et~al.}(2013){Mazeh}, {Nachmani}, {Holczer}, {Fabrycky},
  {Ford}, {Sanchis-Ojeda}, {Sokol}, {Rowe}, {Zucker}, {Agol}, {Carter},
  {Lissauer}, {Quintana}, {Ragozzine}, {Steffen}, \& {Welsh}}]{Mazeh2013}
{Mazeh}, T., {et~al.} 2013, \apjs, 208, 16

\bibitem[{{McQuillan} {et~al.}(2013){McQuillan}, {Mazeh}, \&
  {Aigrain}}]{McQuillan2013}
{McQuillan}, A., {Mazeh}, T., \& {Aigrain}, S. 2013, \apjl, 775, L11

\bibitem[{{Miralda-Escud{\'e}}(2002)}]{Miralda2002}
{Miralda-Escud{\'e}}, J. 2002, \apj, 564, 1019

\bibitem[{{Morley} {et~al.}(2013){Morley}, {Fortney}, {Kempton}, {Marley},
  {Visscher}, \& {Zahnle}}]{Morley2013}
{Morley}, C.~V., {Fortney}, J.~J., {Kempton}, E.~M.-R., {Marley}, M.~S.,
  {Visscher}, C., \& {Zahnle}, K. 2013, \apj, 775, 33

\bibitem[{{Nesvorn{\'y}} {et~al.}(2013){Nesvorn{\'y}}, {Kipping}, {Terrell},
  {Hartman}, {Bakos}, \& {Buchhave}}]{Nesvorny2013}
{Nesvorn{\'y}}, D., {Kipping}, D., {Terrell}, D., {Hartman}, J., {Bakos},
  G.~{\'A}., \& {Buchhave}, L.~A. 2013, \apj, 777, 3

\bibitem[{{Nesvorn{\'y}} {et~al.}(2012){Nesvorn{\'y}}, {Kipping}, {Buchhave},
  {Bakos}, {Hartman}, \& {Schmitt}}]{Nesvorny2012}
{Nesvorn{\'y}}, D., {Kipping}, D.~M., {Buchhave}, L.~A., {Bakos}, G.~{\'A}.,
  {Hartman}, J., \& {Schmitt}, A.~R. 2012, Science, 336, 1133

\bibitem[{{Nesvorn{\'y}} {et~al.}(2014){Nesvorn{\'y}}, {Vokrouhlick{\'y}}, \&
  {Deienno}}]{Nesvorny2014}
{Nesvorn{\'y}}, D., {Vokrouhlick{\'y}}, D., \& {Deienno}, R. 2014, \apj, 784,
  22

\bibitem[{{Ofir} {et~al.}(2014){Ofir}, {Dreizler}, {Zechmeister}, \&
  {Husser}}]{Ofir2014}
{Ofir}, A., {Dreizler}, S., {Zechmeister}, M., \& {Husser}, T.-O. 2014, \aap,
  561, A103

\bibitem[{{Pr{\v s}a} {et~al.}(2011){Pr{\v s}a}, {Batalha}, {Slawson}, {Doyle},
  {Welsh}, {Orosz}, {Seager}, {Rucker}, {Mjaseth}, {Engle}, {Conroy},
  {Jenkins}, {Caldwell}, {Koch}, \& {Borucki}}]{Prsa2011}
{Pr{\v s}a}, A., {et~al.} 2011, \aj, 141, 83

\bibitem[{{Quillen}(2011)}]{Quillen2011}
{Quillen}, A.~C. 2011, MNRAS, 418, 1043

\bibitem[{{Reidemeister} {et~al.}(2009){Reidemeister}, {Krivov}, {Schmidt},
  {Fiedler}, {M{\"u}ller}, {L{\"o}hne}, \& {Neuh{\"a}user}}]{Reidemeister2009}
{Reidemeister}, M., {Krivov}, A.~V., {Schmidt}, T.~O.~B., {Fiedler}, S.,
  {M{\"u}ller}, S., {L{\"o}hne}, T., \& {Neuh{\"a}user}, R. 2009, \aap, 503,
  247

\bibitem[{{Rivera} {et~al.}(2010){Rivera}, {Laughlin}, {Butler}, {Vogt},
  {Haghighipour}, \& {Meschiari}}]{Rivera2010}
{Rivera}, E.~J., {Laughlin}, G., {Butler}, R.~P., {Vogt}, S.~S.,
  {Haghighipour}, N., \& {Meschiari}, S. 2010, \apj, 719, 890

\bibitem[{{Rogers} {et~al.}(2011){Rogers}, {Bodenheimer}, {Lissauer}, \&
  {Seager}}]{Rogers2011}
{Rogers}, L.~A., {Bodenheimer}, P., {Lissauer}, J.~J., \& {Seager}, S. 2011,
  \apj, 738, 59

\bibitem[{{Rowe} {et~al.}(2014){Rowe}, {Bryson}, {Marcy}, {Lissauer},
  {Jontof-Hutter}, {Mullally}, {Gilliland}, {Issacson}, {Ford}, {Howell},
  {Borucki}, {Haas}, {Huber}, {Steffen}, {Thompson}, {Quintana}, {Barclay},
  {Still}, {Fortney}, {Gautier}, {Hunter}, {Caldwell}, {Ciardi}, {Devore},
  {Cochran}, {Jenkins}, {Agol}, {Carter}, \& {Geary}}]{Rowe2014}
{Rowe}, J.~F., {et~al.} 2014, \apj, 784, 45

\bibitem[{{Schmitt} {et~al.}(2014){Schmitt}, {Wang}, {Fischer}, {Jek},
  {Moriarty}, {Boyajian}, {Schwamb}, {Lintott}, {Lynn}, {Smith}, {Parrish},
  {Schawinski}, {Simpson}, {LaCourse}, {Omohundro}, {Winarski}, {Goodman},
  {Jebson}, {Schwengeler}, {Paterson}, {Sejpka}, {Terentev}, {Jacobs},
  {Alsaadi}, {Bailey}, {Ginman}, {Granado}, {Vonstad Guttormsen}, {Mallia},
  {Papillon}, {Rossi}, \& {Socolovsky}}]{Schmitt2014}
{Schmitt}, J.~R., {et~al.} 2014, \aj, 148, 28

\bibitem[{{Schwamb} {et~al.}(2012){Schwamb}, {Lintott}, {Fischer}, {Giguere},
  {Lynn}, {Smith}, {Brewer}, {Parrish}, {Schawinski}, \&
  {Simpson}}]{Schwamb2012}
{Schwamb}, M.~E., {et~al.} 2012, \apj, 754, 129

\bibitem[{{Schwamb} {et~al.}(2013){Schwamb}, {Orosz}, {Carter}, {Welsh},
  {Fischer}, {Torres}, {Howard}, {Crepp}, {Keel}, {Lintott}, {Kaib}, {Terrell},
  {Gagliano}, {Jek}, {Parrish}, {Smith}, {Lynn}, {Simpson}, {Giguere}, \&
  {Schawinski}}]{Schwamb2013}
---. 2013, \apj, 768, 127

\bibitem[{{Seager} \& {Mall{\'e}n-Ornelas}(2003)}]{Seager2003}
{Seager}, S., \& {Mall{\'e}n-Ornelas}, G. 2003, in Astronomical Society of the
  Pacific Conference Series, Vol. 294, Scientific Frontiers in Research on
  Extrasolar Planets, ed. D.~{Deming} \& S.~{Seager}, 419--422

\bibitem[{{Slawson} {et~al.}(2011){Slawson}, {Pr{\v s}a}, {Welsh}, {Orosz},
  {Rucker}, {Batalha}, {Doyle}, {Engle}, {Conroy}, {Coughlin}, {Gregg},
  {Fetherolf}, {Short}, {Windmiller}, {Fabrycky}, {Howell}, {Jenkins}, {Uddin},
  {Mullally}, {Seader}, {Thompson}, {Sanderfer}, {Borucki}, \&
  {Koch}}]{Slawson2011}
{Slawson}, R.~W., {et~al.} 2011, \aj, 142, 160

\bibitem[{{Steffen} {et~al.}(2012){Steffen}, {Fabrycky}, {Ford}, {Carter},
  {D{\'e}sert}, {Fressin}, {Holman}, {Lissauer}, {Moorhead}, {Rowe},
  {Ragozzine}, {Welsh}, {Batalha}, {Borucki}, {Buchhave}, {Bryson}, {Caldwell},
  {Charbonneau}, {Ciardi}, {Cochran}, {Endl}, {Everett}, {Gautier},
  {Gilliland}, {Girouard}, {Jenkins}, {Horch}, {Howell}, {Isaacson}, {Klaus},
  {Koch}, {Latham}, {Li}, {Lucas}, {MacQueen}, {Marcy}, {McCauliff}, {Middour},
  {Morris}, {Mullally}, {Quinn}, {Quintana}, {Shporer}, {Still}, {Tenenbaum},
  {Thompson}, {Twicken}, \& {Van Cleve}}]{Steffen2012}
{Steffen}, J.~H., {et~al.} 2012, \mnras, 421, 2342

\bibitem[{{Tenenbaum} {et~al.}(2013){Tenenbaum}, {Jenkins}, {Seader}, {Burke},
  {Christiansen}, {Rowe}, {Caldwell}, {Clarke}, {Li}, {Quintana}, {Smith},
  {Thompson}, {Twicken}, {Borucki}, {Batalha}, {Cote}, {Haas}, {Hunter},
  {Sanderfer}, {Girouard}, {Hall}, {Ibrahim}, {Klaus}, {McCauliff}, {Middour},
  {Sabale}, {Uddin}, {Wohler}, {Barclay}, \& {Still}}]{Tenenbaum2013}
{Tenenbaum}, P., {et~al.} 2013, \apjs, 206, 5

\bibitem[{{Tenenbaum} {et~al.}(2014){Tenenbaum}, {Jenkins}, {Seader}, {Burke},
  {Christiansen}, {Rowe}, {Caldwell}, {Clarke}, {Coughlin}, {Li}, {Quintana},
  {Smith}, {Thompson}, {Twicken}, {Haas}, {Henze}, {Hunter}, {Sanderfer},
  {Campbell}, {Girouard}, {Klaus}, {McCauliff}, {Middour}, {Sabale}, {Kamal
  Uddin}, {Wohler}, {Barclay}, \& {Still}}]{Tenenbaum2014}
---. 2014, \apjs, 211, 6

\bibitem[{{Torres} {et~al.}(2012){Torres}, {Fischer}, {Sozzetti}, {Buchhave},
  {Winn}, {Holman}, \& {Carter}}]{Torres2012}
{Torres}, G., {Fischer}, D.~A., {Sozzetti}, A., {Buchhave}, L.~A., {Winn},
  J.~N., {Holman}, M.~J., \& {Carter}, J.~A. 2012, \apj, 757, 161

\bibitem[{{Valencia} {et~al.}(2013){Valencia}, {Guillot}, {Parmentier}, \&
  {Freedman}}]{Valencia2013}
{Valencia}, D., {Guillot}, T., {Parmentier}, V., \& {Freedman}, R.~S. 2013,
  \apj, 775, 10

\bibitem[{{Valenti} \& {Fischer}(2005)}]{Valenti2005}
{Valenti}, J.~A., \& {Fischer}, D.~A. 2005, \apjs, 159, 141

\bibitem[{{Valenti} \& {Piskunov}(1996)}]{Valenti1996}
{Valenti}, J.~A., \& {Piskunov}, N. 1996, \aaps, 118, 595

\bibitem[{{Vogt} {et~al.}(1994){Vogt}, {Allen}, {Bigelow}, {Bresee}, {Brown},
  {Cantrall}, {Conrad}, {Couture}, {Delaney}, {Epps}, {Hilyard}, {Hilyard},
  {Horn}, {Jern}, {Kanto}, {Keane}, {Kibrick}, {Lewis}, {Osborne},
  {Pardeilhan}, {Pfister}, {Ricketts}, {Robinson}, {Stover}, {Tucker}, {Ward},
  \& {Wei}}]{Vogt1994}
{Vogt}, S.~S., {et~al.} 1994, in Society of Photo-Optical Instrumentation
  Engineers (SPIE) Conference Series, Vol. 2198, Society of Photo-Optical
  Instrumentation Engineers (SPIE) Conference Series, 362--+

\bibitem[{{Wang} {et~al.}(2013){Wang}, {Fischer}, {Barclay}, {Boyajian},
  {Crepp}, {Schwamb}, {Lintott}, {Jek}, {Smith}, {Parrish}, {Schawinski},
  {Schmitt}, {Giguere}, {Brewer}, {Lynn}, {Simpson}, {Hoekstra}, {Jacobs},
  {LaCourse}, {Schwengeler}, {Chopin}, \& {Herszkowicz}}]{Wang2013a}
{Wang}, J., {et~al.} 2013, \apj, 776, 10

\bibitem[{{Wisdom}(2006)}]{Wisdom2006}
{Wisdom}, J. 2006, \aj, 131, 2294

\bibitem[{{Wisdom} \& {Holman}(1991)}]{Wisdom1991}
{Wisdom}, J., \& {Holman}, M. 1991, \aj, 102, 1528

\bibitem[{{Wolszczan} \& {Frail}(1992)}]{Wolszczan1992}
{Wolszczan}, A., \& {Frail}, D.~A. 1992, \nat, 355, 145

\bibitem[{{Wright} {et~al.}(2011){Wright}, {Fakhouri}, {Marcy}, {Han}, {Feng},
  {Johnson}, {Howard}, {Fischer}, {Valenti}, {Anderson}, \&
  {Piskunov}}]{Wright2011}
{Wright}, J.~T., {et~al.} 2011, \pasp, 123, 412

\bibitem[{{Xie}(2014)}]{Xie2014}
{Xie}, J.-W. 2014, \apjs, 210, 25

\bibitem[{{Zeng} \& {Sasselov}(2013)}]{Zeng2013}
{Zeng}, L., \& {Sasselov}, D. 2013, \pasp, 125, 227

\end{thebibliography}



\clearpage


\LongTables
\begin{deluxetable*}{lcccc}
\tablewidth{0pt}
\tablecaption{Transits of PH3 b, c, and d through January, 2019. \label{tab:transits}}
\tablehead{
\colhead{Planet}      &
\colhead{Transit}     &
\colhead{Midtransit}  &
\colhead{Error}       & 
\colhead{Observed?}   \\ 
\colhead{}            &
\colhead{}            &
\colhead{JD-2454000}  &
\colhead{(days)}      \\ 
}

\startdata

PH3~b   	&	0	&	965.6895	&	0.003	&	  Obs.     \\
PH3~b   	&	1	&	1000.2321	&	0.0028	&	  \nodata  \\
PH3~b   	&	2	&	1034.7782	&	0.0027	&	  Obs.     \\
PH3~b   	&	3	&	1069.3195	&	0.0025	&	  Obs.     \\
PH3~b   	&	4	&	1103.8618	&	0.0023	&	  Obs.     \\
PH3~b   	&	5	&	1138.4057	&	0.0025	&	  Obs.     \\
PH3~b   	&	6	&	1172.9490	&	0.0023	&	  Obs.     \\
PH3~b   	&	7	&	1207.4919	&	0.0024	&	  Obs.     \\
PH3~b   	&	8	&	1242.0336	&	0.0023	&	  Obs.     \\
PH3~b   	&	9	&	1276.5793	&	0.0025	&	  \nodata  \\
PH3~b   	&	10	&	1311.1210	&	0.0024	&	  Obs.     \\
PH3~b   	&	11	&	1345.6650	&	0.0024	&	  Obs.     \\
PH3~b   	&	12	&	1380.2085	&	0.0023	&	  Obs.     \\
PH3~b   	&	13	&	1414.7552	&	0.0023	&	  Obs.     \\
PH3~b   	&	14	&	1449.2979	&	0.0022	&	  Obs.     \\
PH3~b   	&	15	&	1483.8420	&	0.0022	&	  Obs.     \\
PH3~b   	&	16	&	1518.3875	&	0.002	&	  Obs.     \\
PH3~b   	&	17	&	1552.9328	&	0.0021	&	  \nodata  \\
PH3~b   	&	18	&	1587.4768	&	0.0021	&	  Obs.     \\
PH3~b   	&	19	&	1622.0207	&	0.0022	&	  Obs.     \\
PH3~b   	&	20	&	1656.5669	&	0.002	&	  Obs.     \\
PH3~b   	&	21	&	1691.1095	&	0.0021	&	  Obs.     \\
PH3~b   	&	22	&	1725.6528	&	0.0021	&	  Obs.     \\
PH3~b   	&	23	&	1760.1964	&	0.0023	&	  Obs.     \\
PH3~b   	&	24	&	1794.7415	&	0.0022	&	  Obs.     \\
PH3~b   	&	25	&	1829.2833	&	0.0022	&	  Obs.     \\
PH3~b   	&	26	&	1863.8252	&	0.0021	&	  Obs.     \\
PH3~b   	&	27	&	1898.3696	&	0.0024	&	  \nodata  \\
PH3~b   	&	28	&	1932.9122	&	0.0023	&	  Obs.     \\
PH3~b   	&	29	&	1967.4555	&	0.0023	&	  Obs.     \\
PH3~b   	&	30	&	2001.9977	&	0.0024	&	  Obs.     \\
PH3~b   	&	31	&	2036.5440	&	0.0025	&	  Obs.     \\
PH3~b   	&	32	&	2071.0857	&	0.0025	&	  Obs.     \\
PH3~b   	&	33	&	2105.6296	&	0.0026	&	  Obs.     \\
PH3~b   	&	34	&	2140.1743	&	0.0026	&	  Obs.     \\
PH3~b   	&	35	&	2174.7202	&	0.0027	&	  Obs.     \\
PH3~b   	&	36	&	2209.2644	&	0.0028	&	  \nodata  \\
PH3~b   	&	37	&	2243.8076	&	0.003	&	  Obs.     \\
PH3~b   	&	38	&	2278.3550	&	0.003	&	  Obs.     \\
PH3~b   	&	39	&	2312.8982	&	0.0033	&	  \nodata  \\
PH3~b   	&	40	&	2347.4440	&	0.0035	&	  \nodata  \\
PH3~b   	&	41	&	2381.9870	&	0.0037	&	  Obs.     \\
PH3~b   	&	42	&	2416.5347	&	0.004	&	  \nodata  \\
PH3~b   	&	43	&	2451.0763	&	0.004	&	  \nodata  \\
PH3~b   	&	44	&	2485.6201	&	0.0042	&	  \nodata  \\
PH3~b   	&	45	&	2520.1640	&	0.0043	&	  \nodata  \\
PH3~b   	&	46	&	2554.7085	&	0.0045	&	  \nodata  \\
PH3~b   	&	47	&	2589.2511	&	0.0045	&	  \nodata  \\
PH3~b   	&	48	&	2623.7926	&	0.0046	&	  \nodata  \\
PH3~b   	&	49	&	2658.3379	&	0.0046	&	  \nodata  \\
PH3~b   	&	50	&	2692.8794	&	0.0047	&	  \nodata  \\
PH3~b   	&	51	&	2727.4229	&	0.0046	&	  \nodata  \\
PH3~b   	&	52	&	2761.9655	&	0.0047	&	  \nodata  \\
PH3~b   	&	53	&	2796.5116	&	0.0046	&	  \nodata  \\
PH3~b   	&	54	&	2831.0536	&	0.0048	&	  \nodata  \\
PH3~b   	&	55	&	2865.5969	&	0.0048	&	  \nodata  \\
PH3~b   	&	56	&	2900.1420	&	0.0049	&	  \nodata  \\
PH3~b   	&	57	&	2934.6873	&	0.0051	&	  \nodata  \\
PH3~b   	&	58	&	2969.2314	&	0.0052	&	  \nodata  \\
PH3~b   	&	59	&	3003.7751	&	0.0054	&	  \nodata  \\
PH3~b   	&	60	&	3038.3217	&	0.0056	&	  \nodata  \\
PH3~b   	&	61	&	3072.8653	&	0.0058	&	  \nodata  \\
PH3~b   	&	62	&	3107.4093	&	0.006	&	  \nodata  \\
PH3~b   	&	63	&	3141.9539	&	0.0063	&	  \nodata  \\
PH3~b   	&	64	&	3176.4997	&	0.0065	&	  \nodata  \\
PH3~b   	&	65	&	3211.0424	&	0.0067	&	  \nodata  \\
PH3~b   	&	66	&	3245.5846	&	0.0068	&	  \nodata  \\
PH3~b   	&	67	&	3280.1292	&	0.0069	&	  \nodata  \\
PH3~b   	&	68	&	3314.6721	&	0.007	&	  \nodata  \\
PH3~b   	&	69	&	3349.2149	&	0.007	&	  \nodata  \\
PH3~b   	&	70	&	3383.7565	&	0.0071	&	  \nodata  \\
PH3~b   	&	71	&	3418.3018	&	0.007	&	  \nodata  \\
PH3~b   	&	72	&	3452.8429	&	0.0072	&	  \nodata  \\
PH3~b   	&	73	&	3487.3861	&	0.0071	&	  \nodata  \\
PH3~b   	&	74	&	3521.9294	&	0.0072	&	  \nodata  \\
PH3~b   	&	75	&	3556.4751	&	0.0072	&	  \nodata  \\
PH3~b   	&	76	&	3591.0178	&	0.0073	&	  \nodata  \\
PH3~b   	&	77	&	3625.5608	&	0.0074	&	  \nodata  \\
PH3~b   	&	78	&	3660.1070	&	0.0075	&	  \nodata  \\
PH3~b   	&	79	&	3694.6508	&	0.0077	&	  \nodata  \\
PH3~b   	&	80	&	3729.1962	&	0.0078	&	  \nodata  \\
PH3~b   	&	81	&	3763.7393	&	0.0081	&	  \nodata  \\
PH3~b   	&	82	&	3798.2875	&	0.0083	&	  \nodata  \\
PH3~b   	&	83	&	3832.8296	&	0.0085	&	  \nodata  \\
PH3~b   	&	84	&	3867.3748	&	0.0087	&	  \nodata  \\
PH3~b   	&	85	&	3901.9188	&	0.009	&	  \nodata  \\
PH3~b   	&	86	&	3936.4652	&	0.0092	&	  \nodata  \\
PH3~b   	&	87	&	3971.0078	&	0.0093	&	  \nodata  \\
PH3~b   	&	88	&	4005.5504	&	0.0096	&	  \nodata  \\
PH3~b   	&	89	&	4040.0955	&	0.0096	&	  \nodata  \\
PH3~b   	&	90	&	4074.6377	&	0.0098	&	  \nodata  \\
PH3~b   	&	91	&	4109.1811	&	0.0097	&	  \nodata  \\
PH3~b   	&	92	&	4143.7230	&	0.0099	&	  \nodata  \\
PH3~b   	&	93	&	4178.2688	&	0.0098	&	  \nodata  \\
PH3~b   	&	94	&	4212.8099	&	0.0099	&	  \nodata  \\
PH3~b   	&	95	&	4247.3529	&	0.0099	&	  \nodata  \\
PH3~b   	&	96	&	4281.8970	&	0.01	&	  \nodata  \\
PH3~b   	&	97	&	4316.4420	&	0.01	&	  \nodata  \\
PH3~b   	&	98	&	4350.9855	&	0.0101	&	  \nodata  \\
PH3~b   	&	99	&	4385.5284	&	0.0102	&	  \nodata  \\
PH3~b   	&	100	&	4420.0749	&	0.0103	&	  \nodata  \\ \hline
PH3~c   	&	0	&	975.6406	&	0.006	&	  Obs.     \\
PH3~c   	&	1	&	1041.7767	&	0.0041	&	  Obs.     \\
PH3~c   	&	2	&	1107.8402	&	0.0051	&	  Obs.     \\
PH3~c   	&	3	&	1173.9573	&	0.0038	&	  Obs.     \\
PH3~c   	&	4	&	1239.9957	&	0.0048	&	  Obs.     \\
PH3~c   	&	5	&	1306.0651	&	0.0045	&	  Obs.     \\
PH3~c   	&	6	&	1372.0706	&	0.0043	&	  \nodata  \\
PH3~c   	&	7	&	1438.0798	&	0.0048	&	  Obs.     \\
PH3~c   	&	8	&	1504.0555	&	0.0041	&	  Obs.     \\
PH3~c   	&	9	&	1570.0210	&	0.0044	&	  Obs.     \\
PH3~c   	&	10	&	1635.9845	&	0.0039	&	  \nodata  \\
PH3~c   	&	11	&	1701.9489	&	0.0042	&	  Obs.     \\
PH3~c   	&	12	&	1767.9304	&	0.0045	&	  Obs.     \\
PH3~c   	&	13	&	1833.9212	&	0.0051	&	  \nodata  \\
PH3~c   	&	14	&	1899.9437	&	0.0049	&	  \nodata  \\
PH3~c   	&	15	&	1965.9640	&	0.0053	&	  Obs.     \\
PH3~c   	&	16	&	2032.0335	&	0.0052	&	  Obs.     \\
PH3~c   	&	17	&	2098.0799	&	0.005	&	  Obs.     \\
PH3~c   	&	18	&	2164.1877	&	0.0058	&	  Obs.     \\
PH3~c   	&	19	&	2230.2491	&	0.0049	&	  Obs.     \\
PH3~c   	&	20	&	2296.3765	&	0.0061	&	  Obs.     \\
PH3~c   	&	21	&	2362.4398	&	0.0063	&	  Obs.     \\
PH3~c   	&	22	&	2428.5670	&	0.009	&	  \nodata  \\
PH3~c   	&	23	&	2494.6209	&	0.01	&	  \nodata  \\
PH3~c   	&	24	&	2560.7237	&	0.0125	&	  \nodata  \\
PH3~c   	&	25	&	2626.7534	&	0.0125	&	  \nodata  \\
PH3~c   	&	26	&	2692.8036	&	0.0133	&	  \nodata  \\
PH3~c   	&	27	&	2758.7997	&	0.012	&	  \nodata  \\
PH3~c   	&	28	&	2824.7919	&	0.0116	&	  \nodata  \\
PH3~c   	&	29	&	2890.7622	&	0.0099	&	  \nodata  \\
PH3~c   	&	30	&	2956.7226	&	0.0098	&	  \nodata  \\
PH3~c   	&	31	&	3022.6859	&	0.0098	&	  \nodata  \\
PH3~c   	&	32	&	3088.6524	&	0.0108	&	  \nodata  \\
PH3~c   	&	33	&	3154.6410	&	0.012	&	  \nodata  \\
PH3~c   	&	34	&	3220.6401	&	0.0128	&	  \nodata  \\
PH3~c   	&	35	&	3286.6817	&	0.0127	&	  \nodata  \\
PH3~c   	&	36	&	3352.7165	&	0.0133	&	  \nodata  \\
PH3~c   	&	37	&	3418.8067	&	0.0125	&	  \nodata  \\
PH3~c   	&	38	&	3484.8639	&	0.0126	&	  \nodata  \\
PH3~c   	&	39	&	3550.9854	&	0.0125	&	  \nodata  \\
PH3~c   	&	40	&	3617.0521	&	0.0131	&	  \nodata  \\
PH3~c   	&	41	&	3683.1837	&	0.0145	&	  \nodata  \\
PH3~c   	&	42	&	3749.2444	&	0.0154	&	  \nodata  \\
PH3~c   	&	43	&	3815.3622	&	0.0181	&	  \nodata  \\
PH3~c   	&	44	&	3881.4047	&	0.0186	&	  \nodata  \\
PH3~c   	&	45	&	3947.4875	&	0.022	&	  \nodata  \\
PH3~c   	&	46	&	4013.5056	&	0.022	&	  \nodata  \\
PH3~c   	&	47	&	4079.5362	&	0.0233	&	  \nodata  \\
PH3~c   	&	48	&	4145.5248	&	0.0221	&	  \nodata  \\
PH3~c   	&	49	&	4211.5039	&	0.0214	&	  \nodata  \\
PH3~c   	&	50	&	4277.4705	&	0.0194	&	  \nodata  \\
PH3~c   	&	51	&	4343.4309	&	0.0181	&	  \nodata  \\
PH3~c   	&	52	&	4409.3992	&	0.0175	&	  \nodata  \\ \hline
PH3~d   	&	0	&	1069.6618	&	0.0006	&	  Obs.     \\
PH3~d   	&	1	&	1195.5237	&	0.0006	&	  Obs.     \\
PH3~d   	&	2	&	1321.3870	&	0.0006	&	  Obs.     \\
PH3~d   	&	3	&	1447.2535	&	0.0005	&	  Obs.     \\
PH3~d   	&	4	&	1573.1217	&	0.0005	&	  Obs.     \\
PH3~d   	&	5	&	1698.9896	&	0.0004	&	  Obs.     \\
PH3~d   	&	6	&	1824.8571	&	0.0006	&	  Obs.     \\
PH3~d   	&	7	&	1950.7222	&	0.0007	&	  \nodata  \\
PH3~d   	&	8	&	2076.5849	&	0.0006	&	  Obs.     \\
PH3~d   	&	9	&	2202.4468	&	0.0005	&	  Obs.     \\
PH3~d   	&	10	&	2328.3074	&	0.0007	&	  Obs.     \\
PH3~d   	&	11	&	2454.1685	&	0.0011	&	  \nodata  \\
PH3~d   	&	12	&	2580.0310	&	0.0013	&	  \nodata  \\
PH3~d   	&	13	&	2705.8952	&	0.0014	&	  \nodata  \\
PH3~d   	&	14	&	2831.7624	&	0.0014	&	  \nodata  \\
PH3~d   	&	15	&	2957.6309	&	0.0013	&	  \nodata  \\
PH3~d   	&	16	&	3083.4986	&	0.0012	&	  \nodata  \\
PH3~d   	&	17	&	3209.3654	&	0.0013	&	  \nodata  \\
PH3~d   	&	18	&	3335.2298	&	0.0013	&	  \nodata  \\
PH3~d   	&	19	&	3461.0918	&	0.0014	&	  \nodata  \\
PH3~d   	&	20	&	3586.9533	&	0.0017	&	  \nodata  \\
PH3~d   	&	21	&	3712.8141	&	0.0021	&	  \nodata  \\
PH3~d   	&	22	&	3838.6752	&	0.0024	&	  \nodata  \\
PH3~d   	&	23	&	3964.5384	&	0.0026	&	  \nodata  \\
PH3~d   	&	24	&	4090.4036	&	0.0026	&	  \nodata  \\
PH3~d   	&	25	&	4216.2712	&	0.0026	&	  \nodata  \\
PH3~d   	&	26	&	4342.1399	&	0.0025	&	  \nodata  \\

\enddata

\tablecomments{Observed and predicted transit times through January, 2019.}

\end{deluxetable*}


\end{document}